\documentclass[12pt]{iopart}
\usepackage{iopams}  
\expandafter\let\csname equation*\endcsname\relax
\expandafter\let\csname endequation*\endcsname\relax
\usepackage{amsmath}
\usepackage{amssymb}
\usepackage{amscd}
\usepackage{graphicx}%
\usepackage[T1]{fontenc}
\usepackage{color}
\usepackage{colordvi}
\usepackage{xspace}
\usepackage{amsthm}
\usepackage[english]{babel}

\usepackage{xhfill}

\usepackage[utf8]{inputenc}

\usepackage[bookmarks=false,pdfstartview={FitH}]{hyperref}

\usepackage{bbm}

\newtheorem{aX}{Axiom}

\newcommand{\BLUE}[1]{\textcolor{blue}{#1}}

\makeatletter
\renewcommand\tableofcontents{%
  \section*{\contentsname}
    \@starttoc{toc}%
    }
\makeatother

\begin{document}
\selectlanguage{english}

\title[Cooperative pulses in robust quantum control]{Cooperative pulses in robust quantum control: Application to broadband Ramsey-type pulse sequence elements}

\author{Michael Braun}

\address{Department Chemie,
Technische Universit{\"a}t M{\"u}nchen,\\
Lichtenbergstrasse 4, 85747 Garching, Germany}
\ead{braunman@web.de}

\author{Steffen J. Glaser}

\address{Department Chemie,
Technische Universit{\"a}t M{\"u}nchen,\\
Lichtenbergstrasse 4, 85747 Garching, Germany}
\ead{glaser@tum.de}

\vspace{3mm}
\hspace{17mm}{\footnotesize (Dated: April 13, 2014)}\\
\vspace{-3mm}

\pacs{
02.30.Yy, %
03.65.Aa, %
03.67.-a, %
06.20.-f,  %
76.60.-k,  %
76.60.Pc %
}

\begin{abstract}
A general approach is introduced for the efficient simultaneous optimization of pulses that compensate each other' s imperfections
within the same scan.
This is applied to broadband Ramsey-type experiments, resulting in 
pulses with significantly shorter duration compared to individually optimized broadband pulses.
The advantage of the cooperative pulse approach is demonstrated experimentally for the case of two-dimensional nuclear Overhauser enhancement spectroscopy. In addition to the general approach, a symmetry-adapted analysis of the optimization of Ramsey sequences is presented. Furthermore, the numerical results led to the disovery of a
powerful class of pulses with a special symmetry property,
which results in excellent performance in Ramsey-type experiments. A significantly different scaling of pulse sequence performance as a function of pulse duration is found for characteristic pulse families, which is explained in terms of the different numbers of available degrees of freedom in the offset dependence of the associated Euler angles.\end{abstract}

\maketitle

\BLUE{\scriptsize
\tableofcontents
}

\section{Introduction}
Sequences of coherent and well-defined pulses play an important role in
the measurement and control of quantum systems. 
Applications of control pulses include nuclear magnetic resonance (NMR)  and electron spin resonance (ESR)
spectroscopy \cite{Ernst, Jeschke}, magnetic resonance imaging (MRI) \cite{Handbook},  metrology \cite{metrology}, 
quantum information processing  \cite{QIP} and 
atomic, molecular and optical (AMO) physics in general \cite{AMO1, AMO2}.
Typically, pulse sequences are defined in terms of 
ideal pulses with unlimited amplitude and negligible duration (hard pulse limit). 
In practice, ideal pulses can often be approximated by
 rectangular pulses
of finite duration, during which the phase is constant and the amplitude is set to the maximum available value.

However, simple rectangular pulses are only able to excite spins with relatively small detunings
(offset frequencies) that are in the order of the maximum pulse amplitude (expressed in terms of the Rabi frequency of the pulse).
For broadband applications, e.g.\ in NMR, ESR or optical spectroscopy  with a large range of offset frequencies or highly inhomogeneous line widths, 
the performance of simple rectangular pulses is not satisfactory and improved performance can be achieved by using shaped or composite pulses. In addition to offset effects, experimental imperfections such as 
 uncertainties in the flip angle and amplitude and phase transients have to be taken into account.
Composite and shaped pulses can provide significantly improved performance by compensating {\it their own} imperfections \cite{Freeman_1980, Levitt_1986, Levitt_Encyc, shaped_pulses}.
However, the improved performance of composite pulses comes at a cost: the pulses
can be significantly longer compared to simple rectangular pulses with a concomitant increase in relaxation losses during the pulses if relaxation times are comparable to the pulse duration.

Optimal control theory provides efficient numerical algorithms for the optimization of time-optimal pulses \cite{BEBOP_limits} 
or of relaxation-optimized pulses \cite{OC-relax-opt1}.
With optimal control algorithms 
such as the gradient ascent pulse engineering (GRAPE) algorithm  \cite{GRAPE, OC_Simpson, Machnes_2011, Fouquieres_2011},
tens of thousands of pulse sequence parameters can be efficiently optimized. This
makes it possible to design pulses without any bias towards a specific family of pulses and
to explore the physical performance limits 
 as a function of pulse duration \cite{BEBOP_limits, Power-BEBOP, UR_limits}. %
This approach has provided pulses with unprecedented bandwidth and robustness with respect to experimental imperfections. 

Most experiments do not only consist of a single pulse, but of highly orchestrated sequences of pulses that are separated by delays, which are either constant or which are varied in a systematic way \cite{Ernst}. This opens up additional opportunities to 
improve the overall performance of experiments beyond what is achievable by simply combining
the best possible {\it individually} optimized (composite or shaped) pulses.
This makes it possible to
leverage on the interplay within a pulse sequence
and to exploit the potential of the pulses to compensate {\it each other's} imperfections in a given pulse sequence.
The {\it cooperativity} of such pulses provides important additional degrees of freedom in
the pulse sequence optimization because the individual pulses do {\it not} need to be perfect. The analysis and systematic optimization of cooperative effects between different pulses
promises a better overall performance of pulse sequences
and shorter pulse durations.

\begin{figure}[Hb!]
\hskip 2.6cm
\includegraphics[width=.75\textwidth]{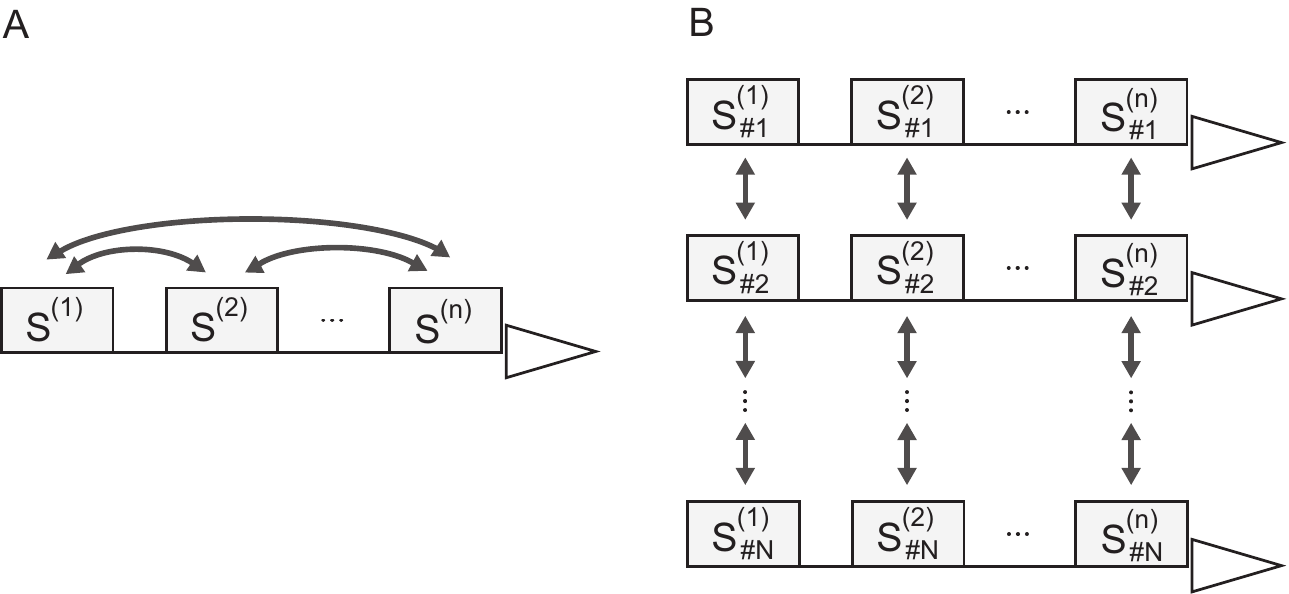} 
\caption{
\small \label{fig:1} 
Two main classes of cooperative pulses are illustrated schematically:
(A) {\it s}ame-{\it s}can {\it coop}erative pulses ({s$^2$-COOP} pulses) and 
(B)  {\it m}ulti-{\it s}can {\it coop}erative pulses ({ms-COOP} pulses).
The rectangles represent (composite or shaped) pulses and the triangles indicate periods of signal acquisition.
The arrows indicate cooperativity between different pulses $S^{(k)}$ and $S^{(l)}$ 
of the {\it same} scan (A) 
or between corresponding pulses $S_{\#i}^{(k)}$ and $S_{\#j}^{(k)}$  
of {\it different} scans $\#i$ and $\#j$ in a multi-scan experiment \cite{COOP_1} (B).
} 
\end{figure}

Here we focus on the analysis and optimization of %
pulse sequences consisting of  individual (composite or shaped) pulses separated by delays.
Together with a final detection period, such a pulse sequence is called a {\it scan}.
Typically, experiments consist of a plurality of scans \cite{Ernst}. In the most simple form of such multi-scan experiments,
a given pulse sequence is simply repeated $N$ times without any modification to accumulate the signal and hence to increase the signal-to-noise ratio.
However, the power of modern coherent spectroscopy results to a large extent
from the systematic variation of the pulses and delays in the different scans,
enabling e.g.\ the suppression of artifacts by phase cycling, the selection of desired coherence transfer pathways, and multi-dimensional spectroscopy or imaging \cite{Ernst}.

In the analysis of cooperativity between pulses, it is useful to distinguish two main classes:
{\it cooperativity} between pulses in the {\it same scan}, i.e.\ between pulses that form a pulse sequence (cf. Fig.\ 1 A) and
{\it cooperativity} between corresponding pulses in {\it different scans} (cf. Fig.\ 1 B).
In order to clearly distinguish these two pulse classes, we propose the terms
{\bf s}ame-{\bf s}can {\bf coop}erative pulses (s$^2$-COOP) for the first class and 
{\bf m}ulti-{\bf s}can {\bf coop}erative pulses (ms-COOP) for the second class.

The mutual cancellation of pulse phase imperfections in the same scan has been 
denoted as {\it global pulse sequence compensation} \cite{Levitt_1986, Levitt_1983, Levitt_Encyc}.
In this approach, a series of ideal hard  pulses is replaced by a series of so-called variable rotation pulses, for which the overall rotation has the same Euler angle $\beta$ but different Euler angles $\gamma$ and $\alpha$ compared to the Euler angle decomposition of the corresponding ideal hard pulses. For a special class of so-called composite LR pulses \cite{Levitt_Encyc},
an explicit procedure was derived to construct a sequence of variable angle rotation pulses that can replace ideal pulses in any given pulse sequence. In addition to constant terms, for LR pulses the $\gamma$ and $\alpha$ angles have a {\it linear} offset dependence of opposite sign, which makes it possible to balance the phase shift created by one pulse by an equal and opposite phase shift  associated with the following pulse. 
Further examples of mutual compensation of offset-dependent phase errors are carefully chosen combinations of 
chirped  pulses  %
\cite{Bodenhausen, Shaka, Garwood_Echo},
which have also been applied to Ramsey-type sequences \cite{Jeschke_Chirp, Suter_Ramsey}.
Results of the simultaneous optimization of excitation and reconversion pulses of a double quantum filter
in solid state NMR have been presented in \cite{OC_Horror},
 but a general approach to s$^2$-COOP pulses has not been analyzed or discussed.

The optimization of cooperativity between corresponding pulses in different scans
has been formulated as an optimal control problem \cite{COOP_1}.
An efficient algorithm was
developed that makes it possible to concurrently optimize a set of ms-COOP pulses.
This algorithm optimizes the 
overall performance of a number of scans,
 leading to an {\it average} signal with desired properties, where undesired terms that may be present in the signal of the individual scans cancel each other.
The class of ms-COOP pulses generalizes the well-known and widely used concepts of phase-cycles \cite{phase_cycle_ref1, phase_cycle_ref2, phase_cycle_ref3}  and difference spectroscopy. The power of this generalization was 
demonstrated both theoretically and experimentally for a variety of applications. %

In this paper, a general {\it filter-based} optimal control algorithm for the simultaneous optimization 
of s$^2$-COOP pulse sequences will be introduced in section 4.1. 
In order to illustrate the approach, a systematic study of cooperativity between 90$^\circ$ pulses in Ramsey-type 
frequency-labeling 
sequences 
\cite{Ramsey1950, Ramsey1957} will be presented.
A special focus will be put on the analysis of the available degrees of freedom and the scaling of overall pulse sequence performance as a function of pulse durations.

\section {The Ramsey scheme}
In his seminal paper published in 1950, Norman Ramsey introduced 
the so-called {\it separated oscillatory fields} method for molecular beam experiments \cite{Ramsey1950},
for which he received the Nobel prize in 1989 \cite{Ramsey_Nobel_lecture}.
He also realized that this approach can be generalized to 
{\it successive oscillatory fields}, i.e.\  
pairs of phase coherent pulses that are not separated in {\it space} but only in {\it time} by a delay $t$ in other experimental settings \cite{Ramsey1957}.
One of the most important early applications of the method was to increase
the accuracy of 
atomic clocks. Even today, most AMO precision measurements 
rely on some variant of the Ramsey scheme \cite{metrology}.
This scheme is also
one of the fundamental experimental building blocks 
in magnetic resonance
and is widely used in 
NMR, ESR and MRI.
For example, the  Ramsey sequence is a key element of stimulated echo experiments \cite{Handbook, Jeschke} 
and 
serves as a frequency-labeling element in many
two-dimensional correlation experiments \cite{Ernst, phase_cycle_ref1, Jeener1979,  Euler4, rectpulse_R}.

\subsection {Objective of Ramsey-type pulse sequences}
In the original paper \cite{Ramsey1950}, the overall effect of the Ramsey scheme was discussed in terms of the created frequency-dependent transition probabilities.
In systems where the initial Bloch vectors are oriented along the $z$-axis, 
the objective of the Ramsey scheme 
can be formulated in terms of  a desired cosine modulation of the $z$-component of the Bloch vector %

\begin{equation}
\label{Mztarget}
{M}^{\rm target}_z(\tau)= s_R\ \cos\{\omega (\tau+\delta)\},
\end{equation}
where $\tau$ is a {\it freely  adjustable} inter-pulse delay that can be chosen by the experimenter
and 
$\delta$ is an optional additional delay that is 
{\it fixed}.
As $\delta$ is the minimum value for the overall effective evolution time 
\begin{equation}
\label{teff0}
t_{\it eff} =\tau 
+\delta,
\end{equation}
in some cases it is desirable to design experiments such that $\delta=0$
and efficient implementations of this condition will be discussed in the following.
However, in many applications of Ramsey-type pulse sequences, the condition $\delta=0$
would pose an unnecessary restriction and the option to allow for $\delta\ne0$ has important consequences for the efficiency and the minimum duration of broadband Ramsey pulses ({\it vide infra}).
In order to obtain the highest contrast (or "visibility") of the Ramsey fringe pattern, the absolute value of the scaling factor $s_R$ should be as large as possible. In the following, we will generally assume $s_R=1$ for ${M}^{\rm target}_z(\tau)$ (but the case of 
$s_R=-1$ will also be considered).
\begin{figure}[Hb!]
\hskip 5.6cm
\includegraphics[width=.3\textwidth]{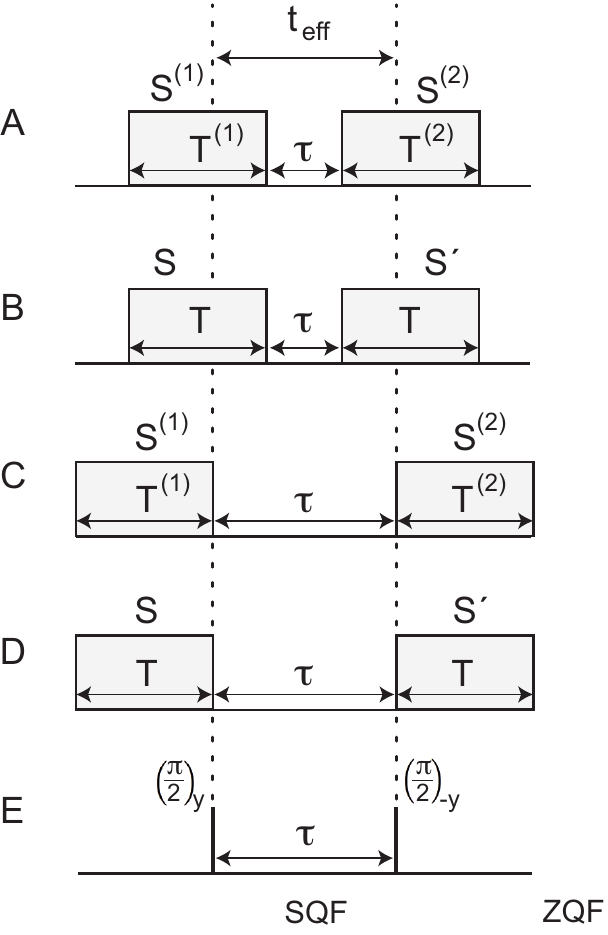} 
\caption{
\small \label{fig:2} 
Characteristic families of Ramsey sequences are schematically represented. Sequences A-D consist of 
(rectangular, shaped or composite) pulses with finite maximum amplitude %
and {\it finite durations} (rectangles).
The inter-pulse delays  are denoted as $\tau$.
The dashed vertical lines separated by the delay $t_{\it eff}=\tau + \delta$ mark the effective evolution time 
(cf.\ Eqs. %
\eqref{teff0} and 
\eqref{teff}) of  the Ramsey sequences.
Sequence E shows the
idealized Ramsey sequence, consisting of two ideal hard 90$^\circ$ %
pulses
with unlimited amplitudes and {\it negligible durations}
and pulse phases $y$ and $-y$. 
After the first and the second pulse of all Ramsey sequences (A-E), an effective single quantum filter (SQF) and a zero quantum filter (ZQF) is applied.
} 
\end{figure}

\begin{figure}[Hb!]
\hskip 3.9cm
\includegraphics[width=.55\textwidth]{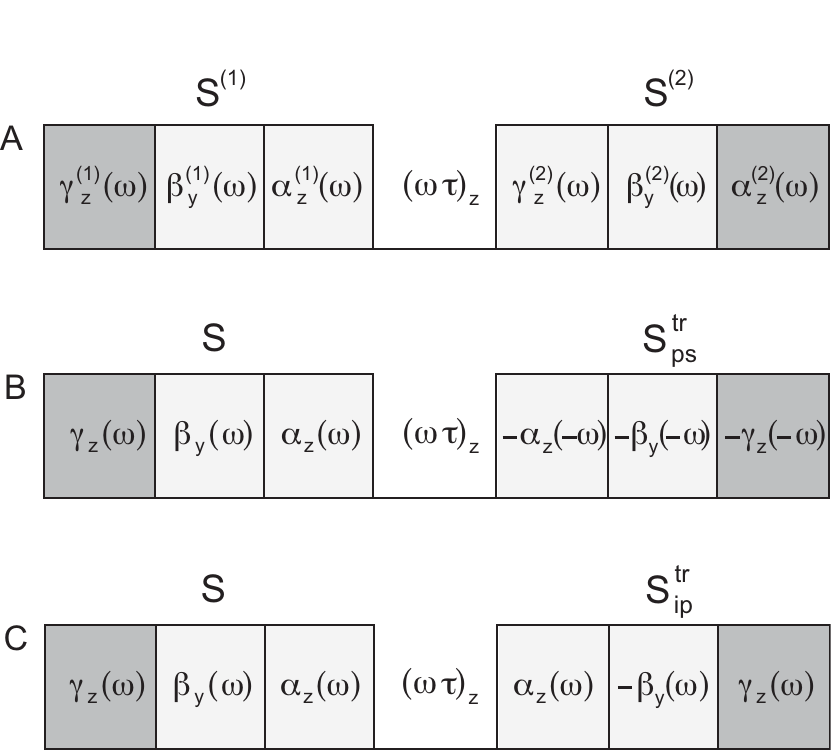} 
\caption{
\small \label{fig:3} 
(A) Schematic representation of the effective rotations of a general Ramsey sequence
$S^{(1)}$-$\tau$-$S^{(2)}$
in terms of the offset-dependent Euler angles of the pulses $S^{(1)}$ and $S^{(2)}$. 
In addition to the
effective Euler rotations $\gamma_z^{(k)}(\omega)$, $\beta_y^{(k)}(\omega)$ and $\alpha_z^{(k)}(\omega)$ of the pulses $S^{(k)}$ with $k \in \{1,2\}$,
the spins are subject to a rotation
by the angle $\omega \tau$ around the $z$-axis 
during the delay $\tau$ between the pulses.
The subscript ($y$ or $z$) of each rotation angle indicates the corresponding rotation axis and $\omega= 2 \pi \nu$ corresponds to the offset in angular frequency units.
In addition, the corresponding transformations for the sequences $S$-$\tau$-$S^\prime$
are shown in (B) for the pulse $S^\prime=S^{\it tr}_{ps}$,
which is a time-reversed version of the pulse $S$ with an additional phase shift by $\pi$,
 and 
in (C) for the pulse
$S^\prime=S^{\it tr}_{ip}$, which is a
time-reversed version of $S$ with inverted phase, cf.\ Table 1. The Euler angles $\gamma_z^{(1)}(\omega)$
and $\alpha_z^{(2)}(\omega)$ are irrelevant for the resulting Ramsey fringe pattern, which is indicated by darker boxes.
} 
\end{figure}

\subsection{Ideal Ramsey pulse sequence}

Fig.\ 2 E shows an idealized hard-pulse version of the Ramsey sequence, consisting of a $90^\circ_y$ pulse, a delay $\tau$ and a $90^\circ_{-y}$ pulse.

The initial Bloch vector is assumed to be oriented along the $z$-axis: 

\begin{equation}
\label{M0}
{\bf M}(0)=%
\left(
0,\ 
0,\ 
1
\right)^{\rm T}
\end{equation}
(The superscript "T" denotes the transpose of the row vector).
The first hard pulse effects an instantaneous $90^\circ$ rotation of negligible duration around the $y$-axis, bringing the Bloch vector
to the $x$-axis of the rotating frame.
During the following delay $\tau$, the Bloch vector rotates around the $z$-axis with the offset  frequency $\omega$, resulting in
\begin{equation}
\label{Mtau1}
{\bf M(\tau)}=
\left(
\cos(\omega \tau),\ 
\sin(\omega \tau),\ 
0
\right)^{\rm T}.
\end{equation}
The second pulse effects an instantaneous $90^\circ$ rotation around the $-y$-axis.
This results in the final Bloch vector
\begin{equation}
\label{Mfinal}
{\bf M^{\it final}}=
\left(
0,\
\sin(\omega \tau),\ 
\cos(\omega \tau)
\right)^{\rm T}
\end{equation}
with the $z$-component
\begin{equation}
\label{Mzfinal}
{M}_z^{\it final}=\rm cos(\omega \tau),
\end{equation}
which in fact has the form of the
target modulation defined in Eq.\ (1) (with $\delta=0$).

\section {Ramsey sequences based on composite pulses with finite amplitude}

In the following, we will use the generic term ``pulse'' for rectangular, composite or shaped pulses. 
Each pulse $S$ is characterized by its duration $T$, 
the time-dependent pulse amplitude $u(t)$ and %
the %
pulse phase $\xi(t)$. %
The pulse  amplitude 
is commonly given in terms of the on-resonance Rabi frequency in units of Hz.
Alternatively, the pulse field can be specified in terms of  its $x$- and $y$-components 
$u_x(t)=
u(t)
\cos \xi(t)$
and
$u_y(t)=
u(t)
\sin \xi(t)$.
Here we analyze a general Ramsey experiment consisting of two pulses $S^{(1)}$ and $S^{(2)}$, separated by a delay $\tau$ (cf. Fig.\ 2 A, C and
3 A).
It is always possible to represent the overall effect of each pulse $S^{(k)}$ (with $k\in \{1,\ 2\}$)
by three Euler rotations
$\gamma_z^{(k)}(\omega)$, $\beta_y^{(k)}(\omega)$, $\alpha_z^{(k)}(\omega)$, 
where the subscripts ($z$ or $y$) denote the (fixed)
rotation axis \cite{%
Euler3, Euler4}.  %

This is illustrated schematically in Fig.\ 3A.
During the delay $\tau$ between the pulses, a spin
with offset $\omega$ experiences an additional rotation by the angle 
$\omega \tau$ around the $z$-axis, which is represented as 
$(\omega \tau)_z$ in Fig.\ 3.
Note that the Euler rotations $\gamma_z^{(1)}(\omega)$ and $\alpha_z^{(2)}(\omega)$ are irrelevant for Ramsey experiments, because the initial Bloch vector ${\bf M}(0)=(0,0,1)^{\rm T}$ 
is invariant under %
$\gamma^{(1)}_z$ 
and because
$M_z^{\it final}$ 
is invariant under
$\alpha^{(2)}_z$. 

The remaining relevant rotations  are $\beta_y^{(1)}(\omega)$ and $\alpha_z^{(1)}(\omega)$ for the first pulse, 
the rotation $(\omega \tau)_z$ for a spin with offset frequency $\omega$ during the delay $\tau$
and  $\gamma_z^{(2)}(\omega)$ and $\beta_y^{(2)}(\omega)$ for the second pulse. In addition to these
rotations, it is common practice in spectroscopy to eliminate any remaining $z$-component of the Bloch vector
after the first pulse by a single quantum filter (SQF), because it would be invariant during the delay $\tau$ and hence cannot contribute to the desired $\tau$ dependence of $M_z^{\it final}$. For example, in 2D NMR spectroscopy, any remaining $z$-component results in unwanted "axial peaks" which can obscure the desired "cross peaks" in the final two-dimensional spectrum \cite{Ernst}.
Similarly, as the experimenter is only interested in the $z$-component of the final magnetization, 
the remaining $x$- or $y$-components of the final Bloch vector can either be ignored or can be
actively eliminated using 
a zero-quantum filter (ZQF) (or a $z$ filter) after the second pulse. In practice, single quantum filters and zero-quantum filters can e.g.\ be realized using phase cycles or so-called "homo-spoil" or "crusher" gradients \cite{Ernst, Handbook, phase_cycle_ref1}.
Hence, the overall sequence of relevant transformations can be summarized schematically as:
\begin{equation}
\label{MSequence}
{\bf M}(0)=%
\left(
\begin{array}{c}
0\\
0\\
1
\end{array}
\right)
\  \overset {\beta^{(1)}_y}{\longrightarrow} \ 
\ \overset {\alpha^{(1)}_z}{\longrightarrow} \
\ \overset {\bf SQF}{\longrightarrow} \
\ \overset {(\omega \tau)_z}{\longrightarrow} \
\ \overset {\gamma^{(2)}_z}{\longrightarrow} \
\  \overset {\beta^{(2)}_y}{\longrightarrow} \ 
\ \overset {\bf ZQF}{\longrightarrow} \
\left(
\begin{array}{c}
0\\
0\\
M_z^{\it final}
\end{array}
\right).
\end{equation}
A straightforward calculation    %
yields the following expression for  $M_z^{\it final}$ as a function of the offset frequency $\omega$, the
delay $\tau$ and the 
Euler angles  $\beta^{(1)}(\omega)$, $\alpha^{(1)}(\omega)$, $\gamma^{(2)}(\omega)$, $\beta^{(2)}(\omega)$:
\begin{equation}
\label{Mzfinseq}
M_z^{\it final}=-\sin \{\beta^{(1)}(\omega)\}\ \sin \{\beta^{(2)}(\omega)\}\ \cos \{
\omega  \tau +\alpha^{(1)} (\omega)+ \gamma^{(2)}(\omega)\}.
\end{equation}
For pulses $S^{(k)}$ with offset-independent Euler angles 
\begin{equation}
\label{betaideal}
\beta_{\it ideal}^{(1)}(\omega) = 90^\circ \ \ \ \ {\rm and } \ \ \ \ \beta_{\it ideal}^{(2)}(\omega) = - 90^\circ, 
\end{equation}
the amplitude of the desired time-dependent cosine modulation (cf. Eq.\ \eqref{Mztarget}) of 
$M_z^{\it final}$ is maximized ($s_R=1$)
and has the form
\begin{equation}
\label{Mzfinalphagamma}
M_z^{\it final}=\cos \{
\omega  \tau +\alpha^{(1)} (\omega)+ \gamma^{(2)}(\omega)\}.
\end{equation}
Furthermore, 
 we can decompose the offset-dependent Euler angles 
$\alpha^{(1)} (\omega)$ and $\gamma^{(2)}(\omega)$ in linear and nonlinear parts in the form
\begin{equation}
\label{gamma1alpha2}
\alpha^{(1)} (\omega)
=
\omega R_\alpha^{(1)} T^{(1)}
+
\alpha^{(1)nl} (\omega)
\ \ \ {\rm and} \ \ \ 
\gamma^{(2)} (\omega)
=
\omega R_\gamma^{(2)} T^{(2)}
+
\gamma^{(2)nl} (\omega),
\end{equation}
with the relative slopes $R_\alpha^{(1)}$ and $R_\gamma^{(2)}$ \cite {Iceberg}
of the linear offset-dependence 
and the
duration $T^{(k)}$ of pulse $S^{(k)}$, i.e.
the nonlinear terms are given by
\begin{equation}
\label{gamma1nonlalpha2nonl}
\alpha^{(1)nl} (\omega)
=
\alpha^{(1)} (\omega)
-
\omega R_\alpha^{(1)} T^{(1)}
\ \ \ {\rm and} \ \ \ 
\gamma^{(2)nl} (\omega)
=
\gamma^{(2)} (\omega)
-
\omega R_\gamma^{(2)} T^{(2)}.
\end{equation}

This allows us to express 
$M_z^{\it final}$ in the form
\begin{equation}
\label{Mzfwwwga}
M_z^{\it final}=\cos \{
\omega  \tau 
+\omega R_\alpha^{(1)} T^{(1)}
+\omega R_\gamma^{(2)} T^{(2)}
+\alpha^{(1)nl} (\omega)
+\gamma^{(2)nl} (\omega)
\}
\end{equation}
$$=\cos \{
\omega  t_{\it eff} 
+\alpha^{(1)nl} (\omega)
+\gamma^{(2)nl} (\omega)
\}
\ \ \  \ \ \ \ \ \ \ \ \ \ \ \ \ \ \ \ \  $$
with the offset-independent {\it effective evolution time}
\begin{equation}
\label{teff}
t_{\it eff} =\tau 
+R_\alpha^{(1)} T^{(1)}
+R_\gamma^{(2)} T^{(2)}.
\end{equation}
Hence,  if %
 the  {\it nonlinear} terms of the Euler angles 
$\alpha^{(1)} (\omega)$ and $\gamma^{(2)} (\omega)$ 
cancel in Eq.\ \eqref{Mzfwwwga} for a pulse pair $S^{(1)}$ and $S^{(2)}$, i.e.\ if the condition
\begin{equation}
\label{g1nlg2nl}
\alpha^{(1)nl} (\omega)
+\gamma^{(2)nl} (\omega)\overset{!}{=}0
\end{equation}
is satisfied,
the modulation of the $z$-component of the final Bloch vector has the desired form of Eq.\ (1): 
\begin{equation}
\label{Mzfcw}
M_z^{\it final}(\tau)=\cos\{\omega(\tau
+\delta)\}
\end{equation}
with the fixed effective delay
\begin{equation}
\label{delta}
\delta=R_\alpha^{(1)} T^{(1)}
+R_\gamma^{(2)} T^{(2)}.
\end{equation}

Note that according to Eq.\ \eqref{g1nlg2nl} it is {\it not} necessary that the nonlinear terms of the {\it individual} Euler angles $\alpha^{(1)} (\omega)$ and $\gamma^{(2)} (\omega)$  are zero.
 Eq.\ \eqref{g1nlg2nl} opens the possibility to design pairs of Ramsey pulses such that the nonlinear
terms $\alpha^{(1)nl} (\omega)$ and   $\gamma^{(2)nl} (\omega)$ cancel each other.
Similarly, if $\delta=0$ is desired for a given application, 
according to
Eq.\ \eqref{delta} it is {\it not} necessary for the individual relative slopes 
$R_\alpha^{(1)}$ and 
$R_\gamma^{(2)} $ to be zero. As phase slopes can be positive or negative \cite{Iceberg},
it is possible to achieve $\delta=0$, even if the individual phase slopes are nonzero. As discussed in the introduction, the mutual compensation of pulse imperfections is expected to result in superior performance of s$^2$-COOP pulses.
In the next section, two equivalent approaches for the design of s$^2$-COOP pulses for Ramsey sequences will be presented.

\section{Optimization of s$^2$-COOP pulses}

\subsection{Filter-based approach to s$^2$-COOP pulse optimization}

Pulse sequences are designed to result in coherence transfer functions  \cite{Ernst} with
a desired dependence on the system parameters such as resonance offsets or coupling constants, and on the delays between the pulses. 
It is important to realize that 
 transfer functions 
 are not only determined by the sequence of pulses but also by inserted {\it filter elements} that are typically realized in practice by 
pulsed field gradients or phase cycles \cite{Ernst, phase_cycle_ref1}.
Depending on the chosen filter elements, the same sequence of pulses
can result in very different transfer functions and hence very different spectroscopic information.
For example, experiments  such as NOESY \cite{Jeener1979, Ernst}, relayed correlation spectroscopy \cite{RELAY} and double-quantum filtered correlation spectroscopy \cite{DQF-COSY} use different  filter elements and
yield very different spectroscopic information although they are all
based on a sequence of three 90$^\circ$ pulses \cite{phase_cycle_ref1}.
If terms of the density operator are filtered based on coherence order \cite{Ernst},
the desired transfer function of a given pulse sequence
is reflected by so-called coherence-order pathways \cite{Ernst}
but more general filter criteria can also be used \cite{ZQF_Keeler, Levitt_Tensor_filters}.

Filters perform non-unitary transformations of the density operator
and in general correspond to projections of the density operator to a subspace of interest.
For example if the Ramsey sequence
is applied to a two-level system, where the state of the system is completely described by the Bloch vector, a single quantum filter (SQF) is numerically simply implemented by
setting the $z$-component of the Bloch vector to zero.
In the GRAPE algorithm, filters can
be treated in complete analogy to relaxation losses \cite{GRAPE}:
In each iteration, 
the Bloch vector ${\bf M}(t)$ (or in general the density operator) evolves forward in time, starting from a given initial state ${\bf M}(0)$ and also
passes the filters forward in time.
For a given final cost (quality factor) $\Phi$, 
the corresponding final costate vector 
\cite{BEBOP1, GRAPE}
\begin{equation}
\label{lambdaf}
{\lambda}_f=%
\left(
\partial\Phi/\partial M_{x}(T_f),\
\partial\Phi/\partial M_y(T_f),\
\partial\Phi/\partial M_z(T_f)
\right)^{\rm T}
\end{equation}
at the final time $T_f$ %
is evolved backward and also
 passes the filters backward in time. 
 (Passing a filter backward in time has the same effect as passing it forward in time,
 e.g.\ a SQF sets the
 $z$-component of the Bloch vector to zero
 in both directions.) 
 The evolution of  $M(t)$ and $\lambda(t)$ is shown schematically in Fig. 4 A.
 With the known state and costate vectors $M(t)$ and $\lambda(t)$, the high-dimensional gradient of the final cost with respect to the 
 control amplitudes can  be efficiently calculated \cite{BEBOP1, GRAPE, Fouquieres_2011} in each iteration step. This gradient information can then be used to update the control parameters in each iteration until convergence is reached. 

This procedure makes it possible to optimize the desired transfer function of an  entire sequence of pulses such that they can compensate each other's imperfections in the best possible way. The full flexibility of the available degrees of freedom is exploited by this approach, resulting in optimal s$^2$-COOP pulses. Most notably, in this approach the number of pulses in a sequence is not limited.

In the case of the Ramsey sequence, for each offset $\omega$, the final figure of merit $\Phi^{(a)}(\omega)$ to be maximized can be defined in terms of the deviation of the $z$-component of the final Bloch vector ${\bf M}^{\it final}(\omega)={\bf M}(T_f, \omega)$ from the target modulation ${\bf M}^{\it target}_z(\omega)$ defined in Eq.\ \eqref{Mztarget}:
\begin{equation}
\label{Phia}
\Phi^{(a)}(\omega)=1-\{M_z^{\it final}(\omega)-M^{\it target}_z(\omega)\}^2
\end{equation}
and according to Eq.\ \eqref{lambdaf}, the final costate vector $\lambda^{(a)}_f$ is given by 
\begin{equation}
\label{lambdafa}
\lambda^{(a)}_f(\omega)
=2
\left(
0,\
0,\
M^{\it target}_z(\omega)-M_z^{\it final}(\omega)
\right)^{\rm T}=2
\left(
0,\
0,\
\cos\{\omega(\tau+\delta)-M_z^{\it final}(\omega)\}
\right)^{\rm T}.
\end{equation}
Formally,   in Eqs.\ \eqref{Mztarget}, \eqref{Mzfinseq} and  \eqref{Mzfcw},
  the inter-pulse delay $\tau$ can also be chosen to be negative. If it is chosen to be $-\delta$ (and assuming $s_R=1$), this results in
  $M_z^{\it target}=1$. 
Hence, %
an alternative figure of merit 
can be defined simply as %
\begin{equation}
\label{Phib}
\Phi^{(b)}(\omega)=M^{\it final}(\omega)\ \  {\rm for} \ \ \tau=-\delta,
\end{equation}
which should be as large as possible and ideally should approach its maximum value of 1 for all offsets $\omega$. In this case, the final costate vector $\lambda^{(b)}_f$ is simply given by 
\begin{equation}
\label{lambdafb}
\lambda^{(b)}_f(\omega)
=
\left(
0,\
0,\
1
\right)^{\rm T}.
\end{equation}
(The two alternative offset-dependent quality factors $\Phi^{(a)}(\omega)$ and $\Phi^{(b)}(\omega)$ 
are closely related to the quality factors for individual pulses used in \cite{tailoringthecostfunction}
and \cite{BEBOP1}, respectively.)

In the GRAPE algorithm \cite{GRAPE}, the overall quality factor $\Phi$ of a given pulse sequence is defined as the average of the local quality factors over the offset range of interest and the gradient for the overall quality factor is simply the average of the offset dependent gradients. In complete analogy to variations in offsets, variations of the scaling factor of the control amplitude can be taken into account \cite{GRAPE}.

\subsection{Symmetry-adapted approach to s$^2$-COOP pulse optimization}

Here an alternative, symmetry-adapted approach for the optimization of s$^2$-COOP 
pulses is introduced.
Although in the case of Ramsey pulses,
this approach is equivalent to the filter-based approach discussed in the previous section,
 it is worthwhile to be considered as it provides a different perspective, which elucidates the inherent symmetry of the problem to design
a pair of maximally cooperative Ramsey pulses.
In order to prepare the detailed discussion of this approach,  we briefly summarize the effects of 
time reversal,
phase inversion and phase shift on the Euler angles associated with a given composite or shaped pulse $S$.

\subsubsection{Effects of time reversal, phase inversion and phase shift by $\pi$}

In order to understand the symmetry-adapted approach to s$^2$-COOP pulses
as well as the
construction principles of Ramsey sequences based on the classes of pulses discussed in sections 6 and 7, it is helpful to consider how the Euler angles $\gamma(\omega)$, $\beta(\omega)$, and $\alpha(\omega)$ of a given pulse $S$  with duration $T$, amplitude $u(t)$ and phase $\xi(t)$
are related to the 
Euler angles $\gamma^\prime(\omega)$, $\beta^\prime(\omega)$, and $\alpha^\prime(\omega)$
of a modified pulse  $S^\prime$. We consider the following 
three symmetry relations \cite{Symm1} (and combinations thereof)
between $S$ and S$^\prime$:

$\bullet$  {\it phase shift} by $\pi$ (denoted as "{\it ps}")

$\bullet$ {\it inversion} of {\it phase} (denoted as "{\it ip}")

$\bullet$ {\it time reversal} of the pulse amplitude and phase (denoted as "{\it tr}")

\noindent The explicit definitions of these pulse modifications in terms of the time-dependent
pulse amplitudes $u(t)$ and phases $\xi(t)$ 
are summarized in the second and third column of Table 1.
In addition to the pulses 
$S_{ps}$, $S_{ip}$ and $S^{tr}$, Table 1 also includes the combinations
  $S_{ps}^{tr}$ and $S_{ip}^{tr}$ (and for completeness the original pulse
  $S$).
 For each of these pulse modifications, the 
relations between the Euler angles of  $S^\prime$ and $S$ are 
summarized in the Table. Explicit derivations of these relations are provided in the Appendix.
In addition, the relation between the Euler angles of a pulse $S$ and its inverse $S^{-1}$ is given in the last row of Table 1. %
Note that for non-zero offset frequencies $\omega$ none of the modified pulses $S_{ps}$, $S_{ip}$, $S^{tr}$, $S_{ps}^{tr}$ and $S_{ip}^{tr}$
 is identical to the inverse $S^{-1}$ of the pulse. Either the order of the Euler angles, their algebraic signs or the sign of the offset frequency where they are evaluated are different.
In particular, it is important to note that simply reversing the amplitude and phase of a pulse $S$ in time (yielding the pulse $S^{tr}$) does {\it not} correspond to $S^{-1}$,
except for the on-resonance case $\omega=0$, where the detuning is zero. Hence 
the "time-resersed" pulse $S^{tr}$ does {\it not} have the same effect as a backward evolution in time.
As shown in Table 1,
the modified pulses $S_{ip}^{tr}$
and 
$S_{ps}^{tr}$
have the closest relation with $S^{-1}$:
The time-reversed pulse with inverted phase $S_{ip}^{tr}$
has the same order of the Euler angles as $S^{-1}$ and they are also evaluated at the same offset frequency $\omega$, the only difference is the algebraic sign of the angles $\gamma$ and $\alpha$. This close relationship of $S_{ip}^{tr}$ and $S^{-1}$ is exploited in the symmetry-adapted analysis of Ramsey s$^2$-COOP pulses in section 4.2.2.

The time-reversed pulse with a phase shift of $\pi$ ($S_{ps}^{tr}$) 
has the same order of the Euler angles and 
they have the same sign as for $S^{-1}$, but they are
evaluated at the {\it negative} offset frequency $-\omega$.
As will be shown below,
$S_{ps}^{tr}$ pulses turn out to play a crucial role in the construction of s$^2$-COOP pulses based on 
a special class of pulses (denoted as $ST$ pulses) that will be introduced and rigorously
defined in section 6.

\begin{table}
\begin{center}
\caption
{\label{tab:Euler_resorted}{
Relations of pulse amplitude $u(t)$, pulse phase $\xi(t)$ and the offset-dependent Euler angles
$\gamma(\omega)$,
   $\beta(\omega)$ and
    $\alpha(\omega)$   
for symmetry-related pulses $S$ and $S^\prime$.
}
}
\begin{tabular}{l c c c c c}
\\
  \hline
$S^\prime$ \ \ \ \ \ \ \ \ \  &
 $u^\prime(t)$  &
  $\xi^\prime(t)$ &
   $\gamma^\prime(\omega)$ &
   $\beta^\prime(\omega)$ &
    $\alpha^\prime(\omega)$   \\
  \hline
$S$ &
$u(t)$ &
$\xi(t)$&
$\gamma(\omega)$&
$\beta(\omega)$&
$\alpha(\omega)$\\
\noalign{\vskip 1em}
$S_{ps}$ &
$u(t)$ &
$\xi(t)+\pi$&
$\gamma(\omega)$&
$-\beta(\omega)$&
$\alpha(\omega)$\\
$S_{ip}$ &
$u(t)$ &
$-\xi(t)$&
$-\gamma(-\omega)$&
$-\beta(-\omega)$&
$-\alpha(-\omega)$\\
$S^{tr}$ &
$u(T-t)$ &
$\xi(T-t)$&
$-\alpha(-\omega)$&
$\beta(-\omega)$&
$-\gamma(-\omega)$\\
\noalign{\vskip 1em}
$S^{tr}_{ps}$ &
$u(T-t)$ &
$\xi(T-t)+\pi$&
$-\alpha(-\omega)$&
$-\beta(-\omega)$&
$-\gamma(-\omega)$\\
$S^{tr}_{ip}$ &
$u(T-t)$ &
$-\xi(T-t)$&
$\alpha(\omega)$&
$-\beta(\omega)$&
$\gamma(\omega)$\\
\noalign{\vskip 1em}
$S^{-1}$ &
- &
- &
$-\alpha(\omega)$&
$-\beta(\omega)$&
$-\gamma(\omega)$\\
  \hline
\end{tabular}
\begin{tabular}{l }
\footnotesize{{\it ps}: phase shifted by $\pi$,  {\it ip}: inverted phase,  {\it tr}: time reversal }
 \ \ \ \ \ \ \ 
 \ \ \ \ \ \ \ 
 \ \ \ \ \ \ \  \ \ \ 
\end{tabular}
\end{center}
\end{table}

\subsubsection{Designing s$^2$-COOP Ramsey pulses by the simultaneous optimization of excitation pulses}

In the Ramsey scheme, the two pulses appear to have quite different tasks. 
A more symmetric picture emerges if the effect of the second pulse is effectively analyzed backward in time.

For a given second Ramsey pulse $S^{(2)}$ (with Euler angles 
$\gamma^{(2)}(\omega)$, $\beta^{(2)}(\omega)$, and $\alpha^{(2)}(\omega)$),
let us consider the pulse $\tilde S^{(2)}$ (with Euler angles 
$\tilde\gamma^{(2)}(\omega)$, $\tilde\beta^{(2)}(\omega)$, and $\tilde\alpha^{(2)}(\omega)$), which we define
as the {\it time-reversed} version of $S^{(2)}$ with {\it inverted phase}:
\begin{equation}
\label{S2}
\tilde S^{(2)}=(S^{(2)})^{tr}_{ip}
\ \ \ {\rm and \ conversely \ \ \ }
S^{(2)}=(\tilde S^{(2)})^{tr}_{ip}.
\end{equation}
Using the relations of the Euler angles between a pulse and its time-reverse and phase-inverted version from Table 1, we find 
\begin{equation}
\label{abg}
\gamma^{(2)}(\omega)=\tilde\alpha^{(2)}(\omega),
  \ \ \ \ \ 
  \beta^{(2)}(\omega)=-\tilde\beta^{(2)}(\omega), 
  \ \ \ \ \ 
  \alpha^{(2)}(\omega)=\tilde\gamma^{(2)}(\omega).
\end{equation}

\begin{figure}[Hb!]
\hskip 3.6cm
\includegraphics[width=.6\textwidth]{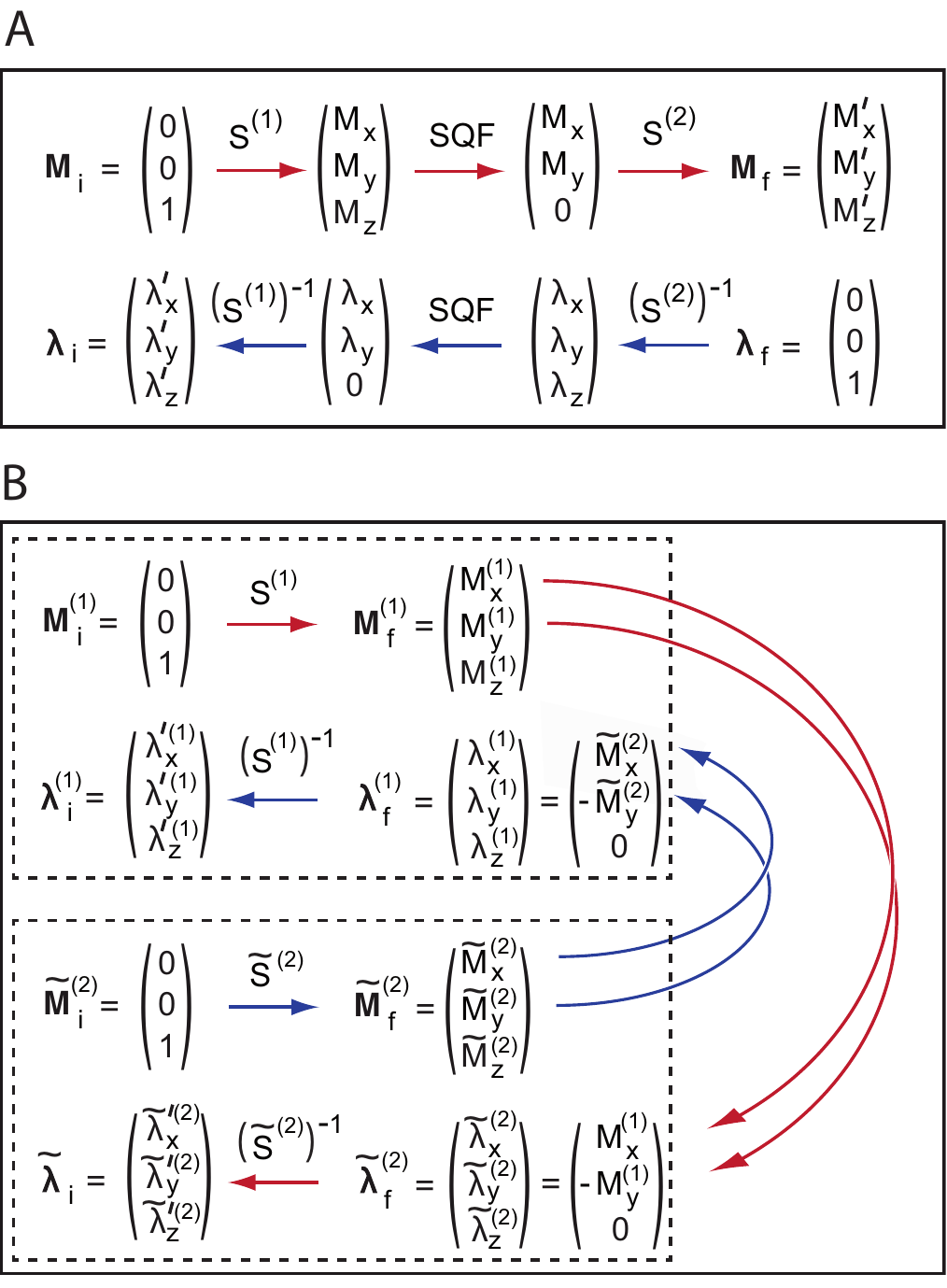} %
\caption{
\small \label{fig:4} 
In panel A, the forward
evolution of  the Bloch vector $M(t)$ and the backward evolution of the costate vector $\lambda(t)$ 
for the filter-based s$^2$-COOP pulse optimization (cf. section 4.1)
are shown for a general Ramsey sequence $S^{(1)}$-$\tau$-$S^{(2)}$. 
Panel B illustrates the evolution of the Bloch and costate vectors
that are considered in the symmetry-adapted approach for the  
simultaneous optimization of the two pulses
$S^{(1)}$ and $\tilde S^{(2)}$, see Eq.\ \eqref{S2} (cf. section 4.2).
}
\end{figure}

\vskip 0.5em
\noindent {\it Special case without auxiliary effective delay} ($\delta=0$):
For simplicity, we first consider the special case where the effective evolution time 
$t_{\it eff}$ is identical to the inter-pulse delay $\tau$, which is the case if the auxiliary fixed effective delay $\delta$ is zero (cf. Fig.\ 2 C).
Based on the general analysis of the Ramsey sequence in section 3 (cf.\ Eqs. \eqref{betaideal} and \eqref{Mzfinalphagamma}),
this case corresponds to the following conditions for the Euler angles of the second pulse:
\begin{equation}
\label{a2b2g2}
\gamma^{(2)}(\omega)=-\alpha^{(1)}(\omega), \ \ \ \ \ \beta^{(2)}(\omega)=-\pi/2, \ \ \ \ \ \alpha^{(2)}(\omega): {\rm arbitrary}.
\end{equation}
Using the relations \eqref{abg}, these conditions translate into the following conditions for 
pulse $\tilde S^{(2)}$:
\begin{equation}
\label{at2bt2gt2}
\tilde \gamma^{(2)}(\omega): {\rm arbitrary} \ \ \ \ \ \tilde \beta^{(2)}(\omega)=\pi/2, \ \ \ \ \ 
\tilde\alpha^{(2)}(\omega)=-\alpha^{(1)}(\omega).
\end{equation}

Hence, the two pulses $S^{(1)}$ and $\tilde S^{(2)}$ play completely symmetric roles.
Both have arbitrary first Euler angles ($\gamma^{(1)}(\omega)$ and  $\tilde \gamma^{(2)}(\omega)$), the second Euler angles ($\beta^{(1)}(\omega)$ and  $\tilde \beta^{(2)}(\omega)$) should both be 90$^\circ$ and the third Euler angles ($\alpha^{(1)}(\omega)$
and $\tilde\alpha^{(2)}(\omega)$) can have an arbitrary offset dependence, provided that for each offset frequency $\omega$ they have the same magnitude but opposite algebraic signs, cf. Eq.\ \eqref{at2bt2gt2}.
The initial %
Bloch vector ${\bf M}_i^{(1)}
=
\left(
0,\
0,\
1
\right)^{\rm T}$
is transferred by the pulse $S^{(1)}$ and a subsequent single quantum filter to
\begin{equation}
\label{Mi1}
{\bf M}_i^{(1)}
\  \overset {\gamma^{(1)}_z}{\longrightarrow} \ 
\  \overset {\beta^{(1)}_y}{\longrightarrow} \ 
\ \overset {\alpha^{(1)}_z}{\longrightarrow} \
\ \overset {\bf SQF}{\longrightarrow} \
{\bf M}_f^{(1)}
=
\left(
\sin {\beta^{(1)}} \cos {\alpha^{(1)}},\
\sin {\beta^{(1)}} \sin {\alpha^{(1)}} , \
0
\right)^{\rm T}.
\end{equation}
Starting from the same initial Bloch vector 
${\bf M}_i^{(2)}={\bf M}_i^{(1)}$,
the pulse $\tilde S^{(2)}$ and a subsequent zero quantum filter yields
\begin{equation}
\label{Mi2tilde}
{\bf \tilde M}^{(2)}_i
\  \overset {\tilde \gamma^{(2)}_z}{\longrightarrow} \ 
\  \overset {\tilde \beta^{(2)}_y}{\longrightarrow} \ 
\ \overset {\tilde\alpha^{(2)}_z}{\longrightarrow} \
\ \overset {\bf SQF}{\longrightarrow} \
{\bf \tilde M}_f^{(2)}
=
\left(
\sin {\tilde \beta^{(2)}} \cos {\tilde \alpha^{(2)}} ,\
\sin {\tilde \beta^{(2)}} \sin {\tilde \alpha^{(2)}} ,\
0
\right)^{\rm T}.
\end{equation}
Ideally, the two final vectors 
${\bf M}_f^{(1)}(\omega)$ 
and
${\bf \tilde M}_f^{(2)}(\omega)$ 
are both located in the transverse plane and have phase angles
${\alpha^{(1)}(\omega)}$
and ${\tilde \alpha^{(2)}(\omega)}$, respectively.
According to the condition
$\tilde\alpha^{(2)}(\omega)=-\alpha^{(1)}(\omega)$ (cf.\ Eq.\ \eqref{at2bt2gt2}),
the two vectors should be collinear 
if one of them is reflected about the $x$-axis. 
Hence, 
a 
figure of merit for the efficiency of the corresponding pair of s$^2$-COOP Ramsey pulses
can be defined as
the scalar product of ${\bf M}_f^{(1)}(\omega)$ 
and the vector resulting from
${\bf \tilde M}_f^{(2)}(\omega)$ if the sign of its $y$-component is inverted:
\begin{equation}
\label{Phicw}
\Phi^{(c)}(\omega)=
\left(
M^{(1)}_{f, x},\
M^{(1)}_{f, y},\
0
\right)
\left(
\tilde M^{(2)}_{f, x},\
-\tilde M^{(2)}_{f, y},\
0
\right)^{\rm T}
=
M^{(1)}_{f, x} \tilde M^{(2)}_{f, x}
-
M^{(1)}_{f, y} \tilde M^{(2)}_{f, y}.
\end{equation}

With this quality factor, the control task to design a sequence of s$^2$-COOP Ramsey pulses
can be formulated as the concurrent optimization of two excitation pulses $S^{(1)}$ and $\tilde S^{(2)}$.
The two identical initial state vectors ${\bf  M}^{(1)}_i$ and 
${\bf \tilde M}^{(2)}_i$  (cf. Eqs. \eqref{Mi1} and \eqref{Mi2tilde})
diverge as they evolve forward in time under the action of the 
two different pulses $S^{(1)}$ and $\tilde S^{(2)}$.
Based on Eqs. \eqref{lambdaf} and \eqref{Phicw}, the two costate vectors associated with
${\bf  M}^{(1)}(t)$ and ${\bf \tilde M}^{(2)}(t)$
are given by
\begin{equation}
\label{lambdaf12}
{\bf  \lambda}^{(1)}_f
=
\left(
\tilde M^{(2)}_{f, x},\
-\tilde M^{(2)}_{f, y},\
0
\right)^{\rm T}
\ \ \ \ {\rm and} \ \ \ \  
{\bf  \tilde \lambda}^{(2)}_f
=
\left(
M^{(1)}_{f, x},\
- M^{(1)}_{f, y},\
0
\right)^{\rm T},
\end{equation}
respectively.
Following the steps of the standard GRAPE algorithm,
the final costate vector
${\bf  \lambda}^{(1)}_f$
is propagated backward in time under the action of the pulse
$S^{(1)}$
and 
${\bf  \lambda}^{(2)}_f$
is propagated backward in time under the action of the pulse
$\tilde S^{(2)}$.
Thus separate gradients of $\Phi^{(c)}(\omega)$ are obtained with respect to
the pulse sequence parameters
of $S^{(1)}$ and $\tilde S^{(2)}$
 \cite{GRAPE, COOP_1}.
Note that the optimizations of 
$S^{(1)}$ and $\tilde S^{(2)}$ are not independent. In fact they are
intimately connected
as according to Eq.\ \eqref{lambdaf12}
the final costate vector after pulse $S^{(1)}$ depends on the $x$ and $y$ coordinates of the final
Bloch vector for pulse 
$\tilde S^{(2)}$ and vice versa.
This is represented graphically in Fig.\ 4 B by the curved arrows.
Fig.\ 4 also illustrates the close
kinship between the optimization of the Ramsey sequence based on
$\Phi^{(b)}$ (cf. Eq.\ \eqref{Phib})
(shown schematically in Fig.\ 4 A)
and the simultaneous optimizations of the excitation pulses
$S^{(1)}$ and $\tilde S^{(2)}$
based on $\Phi^{(c)}$ (cf. Eq.\ \eqref{Phicw})
(Fig.\ 4 B)
for the special case 
$\delta=0$.
Essentially, the evolution of 
${\bf M} (t)$ is folded back after the zero quantum filter ZQF
 between the pulses:
The first part of the entire forward evolution of the Bloch vector ${\bf M} (t)$  in Fig.\ 4 A
(i.e.\ the evolution of ${\bf M} (t)$ under the pulse $S^{(1)}$ up to the 
zero quantum filter ZQF)
corresponds to the
forward evolution of 
${\bf M} ^{(1)}(t)$ under $S^{(1)}$ in Fig.\ 4 B.
The second part of the forward evolution of ${\bf M} (t)$  in Fig.\ 4 A
(i.e.\ the evolution of ${\bf M} (t)$ under the pulse $\tilde S^{(2)}$ starting from the ZQF)
corresponds 
in Fig.\ 4 B
to the
backward evolution of 
${\bf \tilde \lambda} ^{(2)}(t)$ under $\tilde S^{(2)}$.
Similarly, the backward evolution ${\bf \lambda} (t)$ in Fig.\ 4 A 
is folded back in Fig.\ 4 B.
The change of sign of the $y$-components when going from
${\bf M}_f ^{(1)}$ to 
${\bf \tilde \lambda}_f ^{(2)}$ and from
${\bf \tilde M}_f ^{(2)}$
to
${\bf \lambda}_f ^{(1)}$ (cf. curved arrows in Fig.\ 4 B)
is a result
of the construction of the pulse $\tilde S^{(2)}=(S^{(2)})^{tr}_{ip}$. In fact the
two approaches are fully equivalent and the resulting gradients 
are identical.

\vskip 0.5em
\noindent
{\it General case, where the auxiliary effective delay $\delta$ can be non-zero}:
The symmetry-adapted approach outlined above for the special case of  $\delta=0$
can be generalized 
for auxiliary non-vanishing constant effective evolution periods
$\delta$ (cf.\ Fig.\ 2 A).
Whereas for $\delta=0$ the Bloch vectors
${\bf M}_f ^{(1)}$ 
and
${\bf \tilde M}_f ^{(2)}$
should be related by a reflection about the $x$-axis independent of the offset $\omega$,
for $\delta\ne0$ they should be related
by a reflection about an axis in the transverse plane that forms an angle
$(\omega \delta)$
with the $x$-axis.
In this case, 
the quality factor $\Phi^{(c)}(\omega)$
generalizes to
\begin{equation}
\label{Phidw}
\Phi^{(d)}(\omega)
=
\{
M^{(1)}_{f, x} \tilde M^{(2)}_{f, x}
-
M^{(1)}_{f, y} \tilde M^{(2)}_{f, y}
\} \cos (\omega \delta) 
+ 
\{
M^{(1)}_{f, x} \tilde M^{(2)}_{f, y}
+
M^{(1)}_{f, y} \tilde M^{(2)}_{f, x}
\} \sin(\omega \delta).
\end{equation}
and hence in the GRAPE algorithm the corresponding final costate vectors
resulting from Eqs. \eqref{lambdaf}   and \eqref{Phidw}
${\bf  \lambda}^{(1)}_f$
and
${\bf  \tilde \lambda}^{(2)}_f$ are given by
\begin{equation}
\label{lambda1fvec}
{\bf  \lambda}^{(1)}_f
=
\left(
\tilde M^{(2)}_{f, x} \cos (\omega \delta) 
+ \tilde M^{(2)}_{f, y} \sin (\omega \delta),\
\tilde M^{(2)}_{f, x} \sin (\omega \delta)-\tilde M^{(2)}_{f, y} \cos (\omega \delta),\
0
\right)
\end{equation}
and
\begin{equation}
\label{lambda2fvec}
\tilde{\bf  \lambda}^{(2)}_f
=
\left(
M^{(1)}_{f, x} \cos (\omega \delta) 
+ M^{(1)}_{f, y} \sin (\omega \delta) ,\
M^{(1)}_{f, x} \sin (\omega \delta)- M^{(1)}_{f, y} \cos (\omega \delta),\
0
\right).
\end{equation}
As the symmetry-adapted approach to s$^2$-COOP Ramsey pulses yields the optimized pulses
$S^{(1)}$ and $\tilde S^{(2)}$, the pulse
$S^{(1)}$ can be directly used as the first pulse in the Ramsey sequence, whereas the second pulse in the Ramsey sequence is 
$S^{(2)}=(\tilde S^{(2)})^{tr}_{ip}$ (cf. Eq.\ \eqref{S2}).
The optimized s$^2$-COOP pulses $S^{(1)}$ and $S^{(2)}$ presented in the following were developed using the GRAPE algorithm based on
the general Ramsey s$^2$-COOP quality factor
$\Phi^{(d)}(\omega)$.

\section{Examples of s$^2$-COOP Ramsey pulses}

In order to illustrate the power of simultaneously optimized 
cooperative Ramsey pulses
for realistic parameters, 
a challenging problem of practical interest
 was chosen from the field of two-dimensional (2D) NMR spectroscopy.
The 2D NOESY experiment \cite{Jeener1979, Ernst}, which  is widely used to measure 
inter-nuclear distances for the structure elucidation of molecules in solution,
contains a frequency labeling block which is identical to the 
Ramsey sequence. (In the NOESY sequence, the effective evolution time $t_{\it eff}$ between the two Ramsey pulses is called the evolution period $t_1$.)

To facilitate the comparison of the 
relative magnitudes of the
achievable bandwidth of a pulse with
the maximum control amplitude (which is commonly stated in terms of the corresponding Rabi frequency in units of Hz),
in the following we will discuss detunings in terms of the 
offset frequencies 
$\nu=\omega/(2 \pi)$.
As pointed out in the introduction,
simple rectangular pulses cover only a bandwidth $\Delta \nu$
which is in the order of $2 u^{max}$, %
where the bandwidth
 $\Delta \nu=\nu_{max}-\nu_{min}$ is defined as the
difference between the largest and smallest offset frequencies (in units of Hz) with 
acceptable performance. %
For example, in high-resolution $^{13}$C-NMR spectroscopy, the maximum available control amplitude $u^{max}$ is typically in the order of 10 kHz, corresponding to a duration of a rectangular 90$^\circ$ pulse of $T^{(90^\circ)}=1/(4 u^{max})=25\ \mu$s.
With this amplitude,
rectangular pulses only cover a bandwidth of about 20 kHz (corresponding to
maximal or minimal detunings of about $\pm 10$ kHz). %
However, for currently developed NMR spectrometers
with magnetic fields of up to 30 Tesla, %
a bandwidth $\Delta \nu=70$ kHz
will be required in order to cover the typical chemical shift range of 
$^{13}$C spins in proteins. %
Hence, for high-field 
2D $^{13}$C-$^{13}$C-NOESY experiments
\cite{CC_NOESY_1}
 the bandwidth of the Ramsey-type frequency labeling 
 building block should be up to seven times larger than 
$u^{max}$.

For this setting, pulses $S^{(1)}$ and $S^{(2)}$ were optimized simultaneously
using the GRAPE algorithm based on the quality factor
$\Phi^{(d)}$ (Eq.\ \eqref{Phidw}).
In addition to robustness with respect to offset frequencies of up to $\pm 3.5 u^{max}$,
the s$^2$-COOP Ramsey pulses were also optimized to be robust with respect to scaling factors of the control amplitude between 0.95 and 1.05, corresponding to variations of $\pm 5\%$ relative to the nominal control amplitude due to pulse miscalibrations and spatial rf inhomogeneity. %
 For simplicity, both pulses $S^{(1)}$ and $S^{(2)}$ were assumed to have identical pulse durations 
\begin{equation}
\label{T1T2T}
T^{(1)}=T^{(2)}=T.
\end{equation}
The pulses were digitized in steps of 0.5 $\mu$s.
In the optimizations and the subsequent analysis, the overall performance of a given Ramsey sequence is quantified by the total quality factor
\begin{equation}
\label{defphi}
\Phi=\overline{\Phi^{(d)}(\omega)},
\end{equation}
which is the average of $\Phi^{(d)}$ (cf.\ Eq.\ \eqref{Phidw}) over the specified range of offset frequencies and scaling factors of the control amplitude.

\begin{figure}[Hb!]
\centerline{
\includegraphics[width=.5\textwidth]{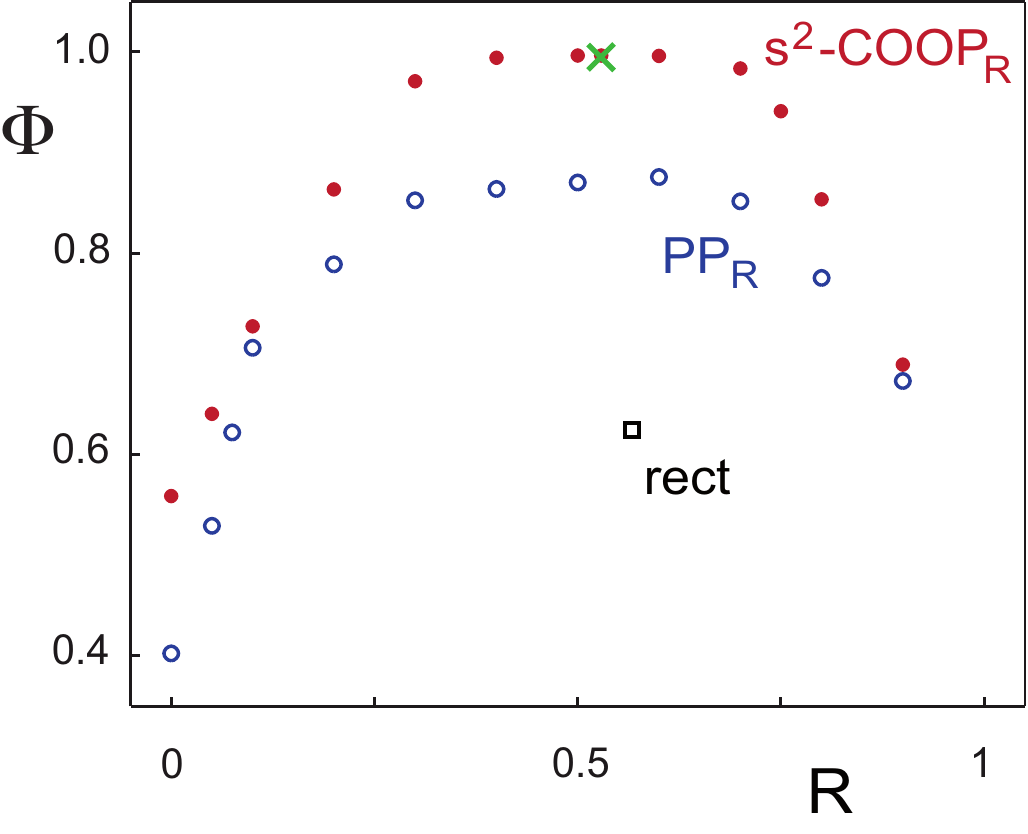} }
\caption{\small \label{fig:5} 
The maximum Ramsey quality factors $\Phi$ (cf.\ Eq.\ \eqref{defphi})
are shown as a function of the 
parameter $R$ (cf.\ Eq.\ \eqref{newr}), which is 
proportional to the
auxiliary  delay $\delta$ (cf.\ Eq.\ \eqref{d2rt}),
for s$^2$-COOP$_R$ pulses (solid circles) and for PP$_R$ pulses (open circles)
with pulse durations of  $T=75\ \mu$s and with a maximum pulse amplitude $u^{max}=10$ kHz.
In addition, the quality factor $\Phi$ is shown for 
Ramsey sequences based on 
the saturation pulse (cross) discussed in section 6.1 
and for
simple rectangular 90$^{\circ}$pulses with a duration of $25$ $\mu$s (open square).
} 
\end{figure}

For fixed pulse durations of  $T=75\ \mu$s (which is only three times longer than the duration of a rectangular $90^\circ$ pulse for $u^{max}=10$ kHz), Fig.\ 5 shows the maximum total quality factors $\Phi$ for s$^2$-COOP Ramsey pulses (solid circles)
as a function of the parameter  
\begin{equation}
\label{newr}
R
={{R_\alpha^{(1)} T^{(1)}+R_{\gamma}^{(2)}T^{(2)}}\over{T^{(1)}+T^{(2)}}}
={{R_\alpha^{(1)}+R_{\gamma}^{(2)}}\over{2}},
\end{equation} 
where the last equality results from Eq.\ \eqref{T1T2T}.
Optimizations were performed for values of $R$ between 0 and 0.9, which are directly related to the durations of  auxiliary constant effective delays
 \begin{equation}
\label{d2rt}
 \delta=2 R T
 \end{equation}   (cf.\ Eq.\ \eqref{delta}).
 The quality factor reaches a plateau with $\Phi\approx1$ between approximately 
 $R=0.3$ and $R=0.7$, corresponding to
 an effective delay $\delta$ between 45 and 105 $\mu$s.
A maximum quality factor 
of $\Phi= 0.9965$
was found for $R=0.53$.
The quality factor decreases markedly as $R$ approaches 0 or 1. For $R=0$, the
auxiliary delay $\delta$ is zero. When $R$ approaches a value of 1,
the effective delay
$\delta$ approaches $2T=150\ \mu$s, i.e.\ it becomes as long as the total duration of the two pulses $S^{(1)}$ and $S^{(2)}$.
For $R\leq 0.6$, all optimized s$^2$-COOP pulses with a duration of $T=75\ \mu$s have a {\it constant} amplitude of $u(t)=u^{max}=10$ kHz, i.e.\ they fully exploit the maximal rf amplitude and  the phase $\xi(t)$ of the pulses is smoothly modulated.

\begin{figure}[Hb!]
\centerline{
\includegraphics[width=.35\textwidth]{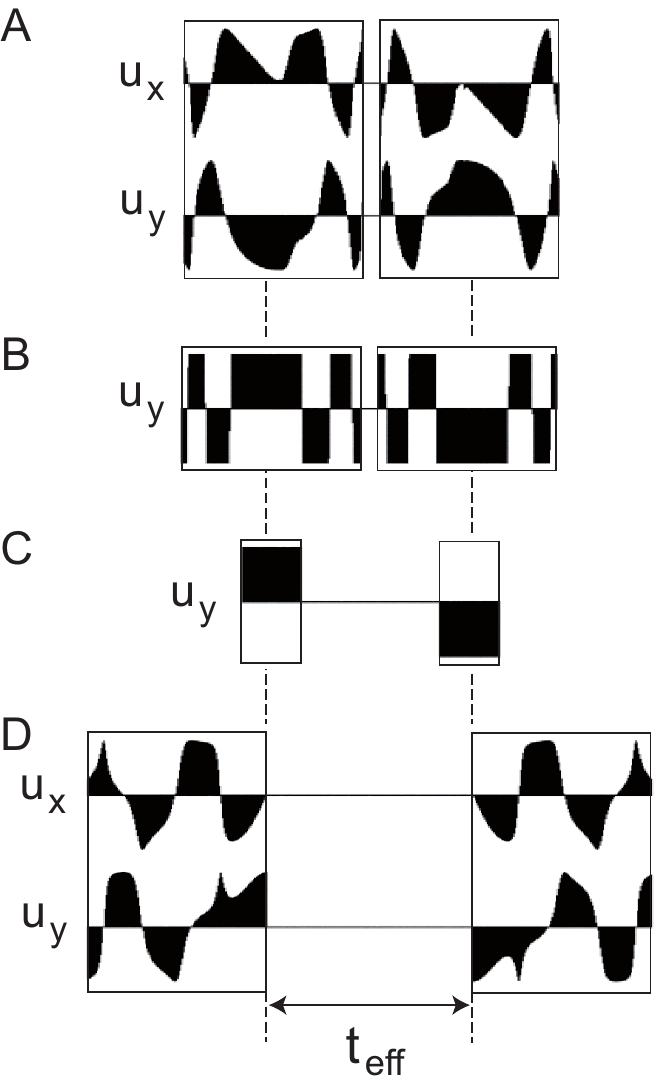} }
\caption{\small \label{fig:6} 
Shapes $u_x(t)$ and $u_y(t)$ 
of optimized
s$^2$-COOP$_R$ (A) and  PP$_R$ (B) pulses with $R=0.53$,
for simple rectangular 90$_{\pm y}^\circ$ pulses (C) and  
PP$_0$ pulses (D). %
For a maximum pulse amplitude $u^{max}$ of 10 kHz,
the duration of the rectangular 90$^\circ$ pulse is 25 $\mu$s (C) and 75 $\mu$s for the remaining pulses (A, B, D). The effective evolution time $t_{\it eff}$ is 85 $\mu$s
and with the auxiliary delays $\delta=79.5$ $\mu$s (A, B), 32 $\mu$s (C) and 0 $\mu$s (D) 
(cf.\ Eq.\ \eqref{d2rt}),  the corresponding
inter-pulse delays $\tau$ are 5.5 $\mu$s (A, B), 53 $\mu$s (C) and 85 $\mu$s, respectively.
} 
\end{figure}

For example, Fig.\ 6 A shows the components $u_x(t)$ and $u_y(t)$ 
of the s$^2$-COOP pulses  $S^{(1)}$ and $S^{(2)}$ for $R=0.53$ (with $\Phi= 0.9965$).
A closer inspection of this figure reveals
an {\it a priori} unexpected symmetry relation between the two pulses:
$u_x^{(2)}$ is a time-reversed copy of $-u_x^{(1)}$ and
$u_y^{(2)}$ is a time-reversed copy of $-u_y^{(1)}$.
Hence the pulse $S^{(2)}$ can be obtained from $S^{(1)}$ by {\it  time-reversal} and an additional {\it phase shift} by $\pi$, i.e.\ $S^{(1)}$ and $S^{(2)}$ have the surprising property
\begin{equation}
\label{S1approxS1trps}
S^{(2)}\approx(S^{(1)})^{tr}_{ps}
\end{equation}
(cf. Table 1).
For the
given parameters and constraints, almost all optimized s$^2$-COOP pulses 
have this special symmetry relation, which was not explicitly imposed in the optimizations.

This finding raises the intriguing question why this particular symmetry relation plays such a prominent role for s$^2$-COOP Ramsey pulse pairs. A detailed analysis  
is given in the next section. This analysis 
led to the discovery and characterization of a powerful new class of pulses (denoted as $ST$ pulses in the following), which
makes it possible to closely approach the excellent performance of 
simultaneously optimized cooperative Ramsey pulses
by a sequence consisting of the pulses $S^{(1)}=S^{ST}$ and  $S^{(2)}=(S^{ST})^{tr}_{ps}$. 
A quantitative comparison of the performance of s$^2$-COOP and ST pulses with the performance of conventional classes of individually optimized pulses in Ramsey sequences will be presented in section 7.

\section{Analysis of Ramsey experiments based on pulse pairs with 
characteristic symmetry relations}

Here we consider constructions
of Ramsey sequence 
$S^{(1)}$-$\tau$-$S^{(2)}$ with 
$S^{(1)}=S$ and $S^{(2)}=S^\prime$,
where 
$S^\prime$ can either be identical to $S$ or can be one of the symmetry related pulses
$S_{ps}$, $S_{ip}$, $S^{tr}$, $S_{ps}^{tr}$ or $S_{ip}^{tr}$
discussed in 
in section 4.2.2. %
As shown in sections 1.2 and  3, 
the desired Ramsey fringe pattern can be expressed in the form of Eq.\ \eqref{Mztarget},
with a
constant auxiliary delay $\delta$,
provided that condition \eqref{g1nlg2nl} is fulfilled, which reduces here to
\begin{equation}
\label{g1nlg2nl1}
\alpha^{nl} (\omega)
+\{\gamma^\prime\}^{nl} (\omega)\overset{!}{=}0,
\end{equation}
where $\alpha^{nl}$ and $\{\gamma^\prime\}^{nl}$
are the nonlinear terms of 
the offset-dependent Euler angles $\alpha$ and
$\gamma^\prime$
of the pulses $S$ and 
$S^\prime$, respectively (cf.\ Eqs.\ \eqref{gamma1alpha2}-\eqref{gamma1nonlalpha2nonl}).

For the sequence $S$-$\tau$-$S^\prime$, the general expression for
$\delta$ 
given in Eq.\ \eqref{delta}
is reduces to
\begin{equation}
\label{delta_ssprime}
\delta
=R_\alpha T + R_{\gamma^\prime} T,
\end{equation}
where
$R_\alpha$ and 
$R_{\gamma^\prime}$
are the relative slopes  of the linear parts of 
the offset-dependent Euler angles $\alpha(\omega)$ and
$\gamma^\prime(\omega)$
of the pulses $S$ and 
$S^\prime$, respectively (cf.\ Eq.\ \eqref{gamma1alpha2}).
With the relations between the Euler angles of $S$ and $S^\prime$ summarized in Table 1, the delay $\delta$ for the sequence 
$S$-$\tau$-$S^\prime$
can be expressed entirely in terms of the relative slopes 
$R_\alpha$ and 
$R_{\gamma}$
and the pulse duration $T$ of pulse $S$. For all considered types of symmetry related pulses $S^\prime$, the resulting expressions for $\delta$
are summarized in the second column of Table 2.
Similarly, using the relations between the Euler angles of $S$ and $S^\prime$ (cf. Table 1),
condition
\eqref{g1nlg2nl1}
can be expressed entirely in terms of the 
nonlinear terms 
$\alpha^{nl}$ and $\gamma^{nl}$ of the Euler angles $\alpha$ and $\gamma$ of the pulse $S$ (cf. Table 2).

\begin{table}
\begin{center}
\label{tab:Constructions_resorted}
\caption
{
Summary of effective delay $\delta$, the Ramsey conditions for the Euler angles $\gamma$ and $\alpha$ of pulse $S$. %
and the scaling factor $s_R$ %
for the Ramsey fringe pattern created by the sequence $S$-$\tau$-$S^\prime$
}
\begin{tabular}{l c c r }
\\
  \hline
$S^\prime$ \ \ \ \ \ \ \ \ \ &
$\delta$&
\ \ \ \ \ \   \ \ \ \ Ramsey condition  \ \   \ \ \ \ \ \ \ \  &
    $s_R$  \\   \hline
$S$ &
$(R_\alpha+R_\gamma)T$ &
$\alpha^{nl}(\omega)\overset{!}{=}-\gamma^{nl}(\omega)$&
$-1$\\
\noalign{\vskip 1em}
$S_{ps}$ &
$(R_\alpha+R_\gamma)T$&
$\alpha^{nl}(\omega)\overset{!}{=}-\gamma^{nl}(\omega)$&
$1$\\
$S_{ip}$ &
$(R_\alpha+R_\gamma)T$&
$\alpha^{nl}(\omega)\overset{!}{=}\gamma^{nl}(-\omega)$&
$1$\\
$S^{tr}$ &
$2 R_\alpha T$ &
$\alpha^{nl}(\omega)\overset{!}{=}\alpha^{nl}(-\omega)$&
$-1$\\
\noalign{\vskip 1em}
$S^{tr}_{ps}$ &
$2 R_\alpha T$ &
$\alpha^{nl}(\omega)\overset{!}{=}\alpha^{nl}(-\omega)$&
$1$\\
$S^{tr}_{ip}$ &
$2 R_\alpha T$&
$\alpha^{nl}(\omega)\overset{!}{=}0 \ \ \ \ \ \ \ \ \ $&
$1$\\
\noalign{\vskip 1em}
$S^{-1}$ &
0&
- \ \ &
$1$\\
  \hline
\end{tabular}
\end{center}
\end{table}

For
the cases $S^\prime=S$ and $S^\prime=S_{ps}$, the Euler angle
$\gamma^\prime(\omega)$ is identical to $\gamma(\omega)$ and 
condition \eqref{g1nlg2nl1}
reduces to $\alpha^{nl} (\omega)
+
\gamma^{nl} (\omega)\overset{!}{=}0$, i.e.\ %
\begin{equation}
\label{g1nlg2nltr}
\alpha^{nl} (\omega)
\overset{!}{=}
-\gamma^{nl} (\omega).
\end{equation}
For the case
$S^\prime=S_{ip}$,
$\gamma^\prime(\omega)=-\gamma(-\omega)$ and 
condition \eqref{g1nlg2nl1}
reduces to $\alpha^{nl} (\omega)
-
\gamma^{nl} (-\omega)\overset{!}{=}0$, i.e.\ %
\begin{equation}
\label{g1nlg2nltr}
\alpha^{nl} (\omega)
\overset{!}{=}
\gamma^{nl} (-\omega).
\end{equation}
Hence for the cases without time reversal, where $S^\prime\in \{S$, $S_{ps}$,
$S_{ip}\}$, condition \eqref{g1nlg2nl1}
corresponds to a condition involving {\it both} Euler angles $\alpha$  {\it and}
$\gamma$. In contrast, for pulses with time reversal, where 
$S^\prime\in \{$$S^{tr}$, $S^{tr}_{ps}$,
$S^{tr}_{ip}\}$, condition \eqref{g1nlg2nl1}
only involves the (nonlinear part of) the Euler angle $\alpha(\omega)$, i.e. the 
performance of the Ramsey sequence $S$-$\tau$-$S^\prime$
is completely independent of $\gamma(\omega)$.
For example, for
$S^\prime=S^{tr}_{ip}$ (cf. Table 1 and Fig. 3 C)
we find 
$\gamma^\prime(\omega)=\alpha(\omega)$ and 
condition \eqref{g1nlg2nl1}
reduces to $\alpha^{nl} (\omega)
+
\alpha^{nl} (\omega)=2 \alpha^{nl} (\omega)\overset{!}{=}0$, i.e.\ %
\begin{equation}
\label{g1nlg2nltr}
\alpha^{nl} (\omega)
\overset{!}{=}
0.
\end{equation}
Hence
for high-fidelity Ramsey sequences with $S^\prime=S^{tr}_{ip}$,
only pulses with a negligible nonlinear offset dependence of the Euler angle $\alpha(\omega)$ are suitable.
This condition is fulfilled by universal rotation (UR) pulses \cite{URconstruction,UR_limits} and point-to-point (PP$_R$) pulses either with $\alpha(\omega)=0$ (corresponding to $R=R_\alpha=0$, cf. Eq.
\eqref{PPEul})
or
so-called Iceberg pulses \cite{Iceberg} with $\alpha(\omega)=R_\alpha T$ (with $R=R_\alpha\ne0$
(cf.\ discussion in section 7.1).

However, a very different situation emerges for Ramsey sequences with
$S^\prime=S^{tr}$ or for $S^\prime=S^{tr}_{ps}$ (cf. Table 1 and Fig. 3 B). In this case,
$\gamma^\prime(\omega)=-\alpha(-\omega)$ and 
condition \eqref{g1nlg2nl1}
reduces to $\alpha^{nl} (\omega)
-
\alpha^{nl} (-\omega)\overset{!}{=}0$, i.e.\ to the condition
\begin{equation}
\label{g1nlg2nltr}
\alpha^{nl} (\omega)
\overset{!}{=}
\alpha^{nl} (-\omega).
\end{equation}
Hence
for $S^\prime=S^{tr}$ and $S^\prime=S^{tr}_{ps}$,
 it is sufficient for
 the nonlinear term of 
$\alpha^{nl} (\omega)$ 
to have a  {\it symmetric} offset dependence
but $\alpha^{nl} (\omega)$ is {\it not} required to be zero.
Only its {\it anti-symmetric} component
\begin{equation}
\label{defgantis}
\alpha^{nl}_a(\omega)={{\alpha^{nl}(\omega)-\alpha^{nl}_s(\omega)}\over {2}}
\end{equation}
has to vanish, i.e.
condition \eqref{g1nlg2nltr} is equivalent to 
\begin{equation}
\label{asas}
\alpha^{nl}_a (\omega)
\overset{!}{=}
0.
\end{equation}
Compared to Eq.\ \eqref{g1nlg2nltr},
this significantly less restrictive condition
suggests that constructions of Ramsey sequences with
$S^\prime=S^{tr}$ or $S^\prime=S^{tr}_{ps}$
offer a decisive advantage compared to other constructions. In fact, this is borne out by the 
results of s$^2$-COOP pulse optimizations without any symmetry constraints presented in section 5,
which are almost exclusively of the form 
$S$-$\tau$-$S^\prime$ with $S^{tr}_{ps}$.

If the conditions specified for each of the 
Ramsey constructions summarized in Table 2
(including $\beta(\omega)=90^\circ$, cf. condition \eqref{betaideal})
are satisfied, %
the sequences create the desired fringe pattern defined in Eq.\ \eqref{Mztarget}. 

Based on Eq.\ \eqref{Mzfinseq} and the relations between $\beta(\omega)$ and 
$\beta^\prime(\omega)$ summarized in Table 1,
it is straightforward to show that the
algebraic sign of the
scaling factor $s_R$
depends on the pulse type $S^\prime$, as indicated in the last column of Table 2.
For the special cases $S^\prime=S$ and $S^\prime=S^{tr}$, the scaling factor is 
$s_R=-1$, whereas for the remaining cases $s_R=1$.
For practical applications, the sign of $s_R$ is irrelevant, as it is always possible to correct for it by simply multiplying the detected signal by $s_R$.
The fact that in section 5 only pulse sequences with
$S^\prime=S^{tr}_{ps}$
but no pulse sequences with $S^\prime=S^{tr}$ were found is a simple consequence of the different algebraic signs of the scaling factor $s_R$ for their Ramsey fringe pattern (cf. Table 2) and the choice
$s_R=1$ for the target  pattern in the optimizations.

Because of its importance, in the next section a novel class of pulses will be formally defined based on 
\eqref{g1nlg2nltr} 
and an efficient algorithm for their direct numerical optimization will be presented.

\subsection{Broadband ST pulses:  {symmetric} offset dependence of the Euler angle $\alpha$ with an optional {tilt}}

We define the class of ST$_R(\beta_0)$ pulses
based on the following three properties of their offset-dependent
Euler angles (cf.\ Table 3).

\vskip 0.5em

\noindent {\it Definition of ST$_R(\beta_0)$ pulses:}

\noindent  (a) It is required that the offset-dependent Euler angle $\alpha(\omega)$ can be expressed in the form
\begin{equation}
\label{defST}
\alpha(\omega)\overset{!}{=}\omega R T+ \alpha_s(\omega),
\end{equation}
where
$\omega R T$ is the part of $\alpha(\omega)$ that is {linear} in $\omega$, $R=R_\alpha$ 
is the relative slope of this linear part  \cite{Iceberg}
and 
\begin{equation}
\label{defgammanlsym}
\alpha_s(\omega)={{\alpha(\omega)+\alpha(-\omega)}\over{2}}
\end{equation}
is the symmetric part of  $\alpha(\omega)$.

\noindent (b) The Euler angle $\beta(\omega)$ is required to have the
offset-independent value \begin{equation}
\label{defbetaST}
\beta(\omega)\overset{!}{=}\beta_0.
\end{equation}

\noindent (c) The Euler angle $\gamma(\omega)$ can have an {\it arbitrary} offset-dependence, i.e. it is {\it not} restricted.

Note that Eq.\ \eqref{defST} in condition (a) is fully
equivalent to  Eqs.\ \eqref{g1nlg2nltr} and  \eqref{asas} (cf. Eq.\ \eqref{gamma1alpha2})
as the 
symmetric part of 
$\alpha(\omega)$
is identical to the 
symmetric nonlinear part of 
$\alpha(\omega)$, 
 because the linear term
$\omega R T$ is anti-symmetric in $\omega$.
Of course for a
finite range of offset frequencies $\omega$,
ST$_R(\beta_0)$ pulses can only be approximated in practice. In the context of Ramsey-type pulse sequences, we focus on the special case where $\beta_0=90^\circ$ (cf. Eq.\ \eqref{betaideal})
and for the sake of simplicity,  we will use the short form
"ST$_R$" (or simply "ST") for
 "ST$_R(90^\circ)$"
  in the following.
For a vanishing relative slope $R=0$, Eq.\ \eqref{defST}  implies that
the offset-dependence of $\alpha(\omega)$
ist strictly symmetric. However, for a non-vanishing relative slope $R\ne0$,
the symmetric 
offset-dependence of $\alpha(\omega)$
is {\it tilted} and the  acronym "ST"  stands for
{\it symmetric} offset dependence of the Euler angle $\alpha$ with an optional {\it tilt}.

\begin{figure}[Hb!]
\centerline{
\includegraphics[width=.9\textwidth]{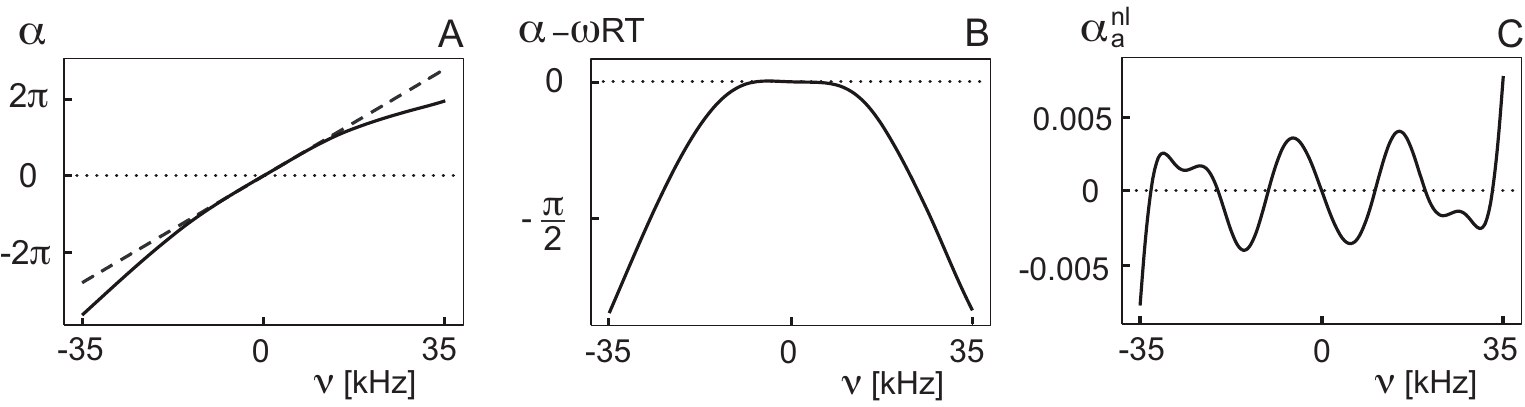} }
\caption{\small \label{fig:7} 
Panel A shows the offset dependence of the Euler angle 
$\alpha(\nu)$ for 
an
optimized ST$_R$ pulse with $R=0.53$.
The dashed line represents the linear part 
 $\omega R T=2 \pi \nu R T$  of $\alpha(\nu)$. %
 Subtracting  $\omega R T$ from 
 $\alpha(\omega)$ 
 yields the
nonlinear part
$ \alpha^{nl}(\nu)$ (cf.\ Eq.\ \eqref{defgammanl}), which is shown in panel B.
Panel C displays the small anti-symmetric nonlinear part $ \alpha_a^{nl}(\nu)$
that remains if
the symmetric nonlinear part 
$\alpha_s^{nl}(\omega)$ (cf.\ Eq.\ \eqref{defgammanlsym})
is subtracted from
$\alpha^{nl}(\omega)$.
} 
\end{figure}

This tilted symmetry is  illustrated in Fig.\ 7 A, which
shows the actual offset dependence of the Euler angle $\alpha(\omega)$ for 
an %
optimized ST$_R$ pulse with $R=0.53$.
The dashed line in Fig.\ 7 A represents the linear part 
 $\omega R T$  of $\alpha(\omega)$. %
Fig.\ 7 B shows the highly symmetric offset dependence of 
the remaining nonlinear part 
\begin{equation}
\label{defgammanl}
\alpha^{nl}(\omega)=\alpha(\omega)- \omega R T
\end{equation}
of $\alpha(\omega)$
(cf.\ Eqs. \eqref{gamma1nonlalpha2nonl}).
For an ideal ST$_R$ pulse, $\alpha^{nl}(\omega)$
should be perfectly symmetric, i.e. the
anti-symmetric nonlinear part $\alpha^{nl}_a(\omega)$
(cf.\ Eq.\ \eqref{defgantis})
should ideally be zero.
This condition is not strictly satisfied
for $\alpha^{nl}_a(\omega)$ in Fig.\ 7 C, 
but it is closely approximated. Within the 
desired range of offsets ($-35$ kHz $\leq \nu=\omega/(2 \pi) \leq 35$ kHz) the largest value of 
$\vert \alpha^{nl}_a(\omega)\vert$ is in the order of 0.005 rad. This corresponds to a maximum absolute deviation of less than 0.3$^\circ$ (corresponding
to a relative error of only about $4 \cdot 10^{-4}$
compared to the largest value of $\vert \alpha(\omega)\vert$ in the offset range of interest).

It is interesting to note that
a so-called saturation pulse with a duration of 75 $\mu$s, that was simply optimized
to bring the Bloch vector from the $z$-axis to the transverse plane by maximizing
the quality factor 
\begin{equation}
\label{phisat}
\Phi^{sat}=M^2_x(T)+M^2_y(T)=1-M^2_z(T)
\end{equation}
\cite{ESR_OC} 
for the same bandwidth of 70 kHz
was also
found to perform surprisingly 
well in Ramsey sequence of the form $S$-$\tau$-$S^{\it tr}_{\it ps}$.
In fact, the Euler angles of this saturation pulse closely approach the 
condition of Eq.\ \eqref{defST} for an ST$_R$ pulse with $R=0.53$, although
 only the desired value of the Euler angle $\beta(\omega) =90^\circ$ was specified, whereas
both $\alpha(\omega)$ and $\gamma(\omega)$ were {\it  a priori} not restricted.
The overall Ramsey quality factor  $\Phi=0.9937$ (cf. Eq.\ \eqref{defphi})  is indicated by a cross in Fig.\ 5 and approaches the quality factor
$\Phi=0.9965$ of
s$^2$-COOP pulses for the same value of $R$. 

The fact that the optimization of a saturation pulse serendipitously resulted in excellent ST$_R$ pulse %
suggests that 
at least for the given optimization parameters (offset range, maximum pulse amplitude, robustness to pulse scaling, pulse duration $T$ etc.) defined in section 5,
ST$_R$ pulses are quite "natural" and can be as short as saturation pulses.
This can be rationalized by considering the Taylor series expansion of
$\alpha(\omega)$:
\begin{equation}
\label{gamtayl}
\alpha(\omega)
=
\alpha^{[0]} (\omega)+\alpha^{[1]} (\omega)+
\alpha^{[2]} (\omega)+\alpha^{[3]} (\omega)+...
\end{equation}
All terms of even order
are symmetric and hence can be arbitrary for ST$_R$ pulses according to condition
\eqref{defST}.
The first-order term in $\omega$
and is given by $\alpha^{[1]} (\omega)=\omega R T$. %
Therefore, the term of lowest order 
that is required to be zero in the Taylor series of $\alpha(\omega)$ in Eq.\ \eqref{gamtayl} is the third-order term $\alpha^{[3]} (\omega)$.
Hence, saturation pulses for which 
$\alpha(\omega)$ can be closely approximated by the lowest order terms of a Taylor series (with order $n< 3$) also
automatically satisfy the conditions for ST pulses.
However,  it is not necessarily the case that %
odd higher-order terms
(with order $n\geq 3$)
can be neglected.
In fact, optimizations of saturation pulses for other optimization parameters 
yielded excellent saturation pulses, which were however poor ST$_R$ pulses (and with poor performance in Ramsey-type experiments).
Therefore, rather than relying on serendipity, it is desirable to have an algorithm for the specific optimization of PS pulses, which will briefly be sketched in the next section.

\subsection{Optimization of individual ST pulses}

In order to study the performance of ST$_R$ pulses in Ramsey-type experiments,
we implemented a straightforward algorithm that makes it possible to
directly optimize 
individual ST pulses based on the conditions defined in Eqs. \eqref{defST}
and \eqref{defbetaST}.
In the version of the  algorithm briefly outlined in the following,
the relative slope $R$ is not fixed but is dynamically
adapted in the iterative procedure in order
 to find the best value of $R$ for the maximum  achievable performance of an ST pulse for
 a given set of optimization parameters.%

The optimization starts with a random pulse shape, which is iteratively refined by the 
optimization algorithm. In each iteration step, 
for a discretized set of offset frequencies $\omega_j$
(with spacing $\omega_{j+1}-\omega_{j}\ll 1/T$), the 
final Bloch vector ${\bf M}(T)$ is calculated
for ${\bf M}(0)=(0,\ 0,\ 1)^{\rm T}$.
The phase of ${\bf M}(T)$ corresponds to $\alpha(\omega)$.
The anti-symmetric part 
\begin{equation}
\label{defgantisym}
\alpha_a(\omega)={{\alpha(\omega)-\alpha(-\omega)}\over{2}}
\end{equation}
of 
$\alpha(\omega)$ is calculated
and unwrapped by adding (or subtracting)  $2\pi$ to (or from) $\alpha_a(\omega_{j+1})$ if $\alpha_a(\omega_{j+1})-\alpha_a(\omega_{j})$ is smaller (or larger) than $-\pi$ (or $\pi$), repectively. From the unwrapped function $\alpha_a(\omega)$, the slope of the extracted linear component $\omega R T$ can be extracted by linear regression.
The symmetric part
$\alpha_s(\omega)$ is calculated using Eq.
 \eqref{defgammanlsym}
 and added to 
the linear component $\omega R T$ to 
define the offset-dependent
 target phase %
\begin{equation}
\label{gamtarg}
\alpha^{\it target}(\omega)= \omega RT + \alpha_s(\omega),
\end{equation}
which is most closely approached by 
 $\alpha(\omega)$ in the current iteration.
 The corresponding target Bloch vector 
\begin{equation}
\label{Mtarg1}
{\bf M}^{\it target}(\omega)=
\left(
\cos\{\alpha^{\it target}(\omega)\},\
\sin\{\alpha^{\it target}(\omega)\},\
0
\right)^{\rm T}
\end{equation}
for the current iteration step is constructed and the
quality factor to be optimized is defined as 
\begin{equation}
\label{qst}
\Phi^{ST}=1-(M_x(T)-M^{\it target}_x)^2-(M_y(T)-M^{\it target}_y)^2-(M_z(T)-M^{\it target}_z)^2.
\end{equation}
The gradient of $\Phi^{ST}$ is calculated according to the standard GRAPE approach \cite{GRAPE, tailoringthecostfunction} and used to update the pulse shape. This updated pulse is then used in the next iteration to re-calculate $\alpha^{\it target}(\omega)$ and the process is repeated until convergence is reached.

For the optimization parameters defined in section 5,
the quality factor $\Phi$ (cf. Eq.\ \eqref{defphi}) of ST$_R$ pulses closely approaches the quality factor of
s$^2$-COOP pulses. For example, the optimization of an ST$_R$ pulse 
resulted in an optimal relative slope of
$R=0.53$ with the quality factor
$\Phi=0.9963$ compared to
$\Phi=0.9965$ for s$^2$-COOP pulses that were directly optimized for $\Phi$
with the same value of $R$.
(Alternatively, ST$_R$ pulses with a desired {\it fixed} relative slope $R$
can be optimized 
by using this fixed value of $R$ in the definition
of 
$\alpha^{\it target}(\omega)$ in \eqref{gamtarg}.)

Based on the analysis and the results presented in sections 5 and 6,
the performance of 
s$^2$-COOP pulses in Ramsey experiments
can be closely approached 
by 
the construction
$S$-$\tau$-$S^{tr}_{ps}$ (or by the construction $S$-$\tau$-$S^{tr}$ with a scaling factor $s_R=-1$, cf. Table 2), provided that the pulse
$S$ satisfies the criteria of an ST$_R$ pulse as defined in section 6.1.
It is interesting to compare the performance of 
s$^2$-COOP and ST-based Ramsey sequences with constructions 
based on established pulse classes.

\section{Comparison of s$^2$-COOP and ST-based Ramsey sequences with constructions based on
conventional pulse classes}

As discussed in section 3, the general Ramsey sequence $S^{(1)}$-$\tau$-$S^{(2)}$
consists of two pulses which in general can be different and don't even need to have the same durations. In addition, in section 6, a number of important Ramsey sequence constructions of the form $S$-$\tau$-$S^\prime$ were discussed, where
$S^\prime \in \{S$, 
$S_{ps}$, $S_{ip}$, $S^{tr}$, $S_{ps}^{tr}$, $S_{ip}^{tr}  \}$ is related to $S$ by simple symmetry relations
(cf. Tables 1 and 2).
In this section, we discuss the associated degrees of freedom (summarized in Table 3) and compare the performance of 
 s$^2$-COOP pulses with 
constructions based on the class of
ST$_R$ pulses introduced in section 6.1
and the well known classes of PP$_R$ and UR pulses.

\subsection{Conventional pulse classes suitable for broadband Ramsey experiments}

In order to be suitable for broadband Ramsey-type experiments,
{\it all} pulses discussed in the following are required to have an 
offset-independent Euler angle 
\begin{equation}
\label{offsunabhbeta}
\beta(\omega)\overset{!}{=}90^\circ
\end{equation}
(cf. Eq.\ \eqref{betaideal}), i.e. they can all be broadly termed $90^\circ$ pulses. Therefore, in the following and in the summary presented in Table 3, we focus on the constraints for the Euler angles $\gamma(\omega)$ and 
$\alpha(\omega)$ that are characteristic for different classes of $90^\circ$ pulses.

\vskip 1em
(a) {\it Broadband UR(90$_y^{\hskip 0.1em\circ}$) pulses}:
Broadband universal rotation (UR) pulses \cite{UR_limits} are designed to effect a rotation with  defined rotation axis
and rotation angle for all offset frequencies $\omega$ of interest. 
Other terms that have been used for
UR pulses are class A pulses
\cite{Levitt_1986}, constant rotation pulses \cite{Levitt_Encyc},
general rotation pulses
\cite{Geen91}, plane rotation pulses \cite{Garwood-BIR} and
universal pulses \cite{Emsley-92}.
For UR(90$_y^\circ$) pulses corresponding to a 90$^\circ$ rotation around the $y$-axis, 
the desired offset-independent Euler angles $\gamma(\omega)$ and $\alpha(\omega)$ are  
\begin{equation}
\label{UREuler}
\gamma(\omega)\overset{!}{=}0, \ \ \ {\rm and} \ \ \   
\alpha(\omega)\overset{!}{=}0,
\end{equation}
as indicated in Table 3.
As conditions \eqref{UREuler} 
imply $R_\alpha=R_\gamma=0$
as well as $\alpha^{nl}(\omega)=\alpha^{nl}_a(\omega)=0$
and
$\gamma^{nl}(\omega)=0$,
the conditions for Ramsey sequences $S$-$\tau$-$S^\prime$ 
summarized in Table 2
are satisfied for all
$S^\prime\in\{S$, 
$S_{ps}$, $S_{ip}$, $S^{tr}$, $S_{ps}^{tr}$, $S_{ip}^{tr}  \}$
and 
the auxiliary delay $\delta$ (cf. Table 2) is always zero for UR $90^\circ_y$
pulses:
\begin{equation}
\label{URdelta}
\delta(UR)=0,
\end{equation}
cf. Fig.\ 2 D.

\vskip 1em
(b) {\it Broadband PP$_R$($z\to x$) pulses}:
In general, point-to-point (PP$_R$) pulses \cite {BEBOP_limits, Iceberg} are designed to transfer a specific initial state to a specific target state.
In the literature, PP$_0$($z\to x$) pulse have also been denoted as
class B2 pulses
\cite{Levitt_1986}.   %
PP$_R$($z\to x$) pulses have been denoted as Iceberg pulses \cite{Iceberg}. %
For the general class of PP$_R$($z\rightarrow x$) pulses considered here, the initial state corresponds to a Bloch vector pointing along the $z$-axis
and for $\omega=0$ (on-resonance case), the target state
corresponds to a vector pointing along the $x$-axis.
For off-resonant spins with $\omega\ne0$, the desired phase 
 of the target vector is a linear function of  $\omega$.
As the initial state is invariant under $z$ rotations, the first Euler angle $\gamma(\omega)$ is irrelevant and can have arbitrary values,  in contrast to the case of universal rotation pulses discussed above.
The second Euler angle has to be $\beta(\omega)=90^\circ$ for all offsets in order to rotate the Bloch vector into the transverse plane by a rotation around the $y$-axis.  This brings the Bloch vector  to the $x$-axis and finally the Euler angle  $\alpha(\omega)$ rotates the Bloch vector in the transverse plane to the desired position with phase $R T \omega$
(with the dimensionless proportionality factor
$R=R_\alpha$ (cf. \cite{Iceberg}) and
$T$ the duration of the pulse). %
Hence, the 
 Euler angle 
 $\alpha(\omega)$
 of a PP$_R$($z\to x$) pulse has to be of the form
 \begin{equation}
\label{PPEul}
\alpha(\omega)\overset{!}{=}R_\alpha T \omega,%
\end{equation}
which
implies that $\alpha^{nl}(\omega)$ and $\alpha^{nl}_a(\omega)$ have to be zero (cf.\ Table 3).
Therefore, the conditions for Ramsey sequences $S$-$\tau$-$S^\prime$
summarized in Table 2
are satisfied for 
$S^\prime\in\{S^{tr}$, $S_{ps}^{tr}$, $S_{ip}^{tr}  \}$,
i.e. for 
all pulses $S^\prime$ for which the relation between
$S$ and $S^\prime$ includes a 
time-reversal operation.
According to Table 2, for these pulses
the auxiliary constant delay $\delta$ is given by
$2 R_\alpha T$.
Hence, for applications where
$\delta$ is required to be zero, also $R_\alpha$ has to be zero:
  \begin{equation}
\label{PPrdel}
R_\alpha  \overset{!}{=}    0 \ \ \ {\rm for} \ \ \  \delta=0,
\end{equation}
cf.\ Fig.\ 2 D.
Conversely, $R_\alpha$ does not have to be identical to zero
in applications where the auxiliary delay $\delta$ is allowed to be non-zero (see Table 3), cf.\ Fig.\ 2 B. Note that $R_\alpha$ is not necessarily positive but can also have negative values \cite{Iceberg, delayed_focus}.
A large pool of highly-optimized broadband
PP$_0$($z\rightarrow x$) pulses  \cite{Levitt_1986, Levitt_Encyc, BEBOP_limits, Croasmun}
and PP$_R$($z\rightarrow x$) pulses \cite{Iceberg}
are available in the literature.

\vskip 1em
(c) {\it Broadband ST$_R(90^\circ)$ pulses}:
For the class of ST pulses
defined in section 6.1,
the nonlinear component of $\alpha(\omega)$
is required to have a vanishing anti-symmetric part
(cf.\ Eqs. \eqref{asas} and %
 Table 3).
Therefore, the conditions for Ramsey sequences $S$-$\tau$-$S^\prime$
summarized in Table 2
are satisfied only for 
$S^\prime\in\{S^{tr}$, $S_{ps}^{tr} \}$.
According to Table 2, for these pulses 
$S^\prime$ the auxiliary constant delay is given by
$2 R_\alpha T$.
Hence, for applications where
$\delta$ is required to be zero, also $R_\alpha$ has to be zero:
  \begin{equation}
\label{PPrdel}
R_\alpha  \overset{!}{=}    0 \ \ \ {\rm for} \ \ \  \delta=0.
\end{equation}

(d) {\it Pulses with Euler angle $\beta=90^\circ$ and arbitrary Euler angles $\alpha(\omega)$ and 
$\gamma(\omega)$}:
In the most general class of pulses that play a role in the context of Ramsey-type experiments, only the desired value of the second Euler angle is fixed ($\beta(\omega)=90^\circ$), whereas the functional form of $\gamma(\omega)$ and  $\alpha(\omega)$ is not restricted.
As discussed in section 6.1,  pulses with this property are called
saturation pulses. They also have been denoted as class B3 pulses \cite{Levitt_1986} and variable rotation pulses \cite{Levitt_Encyc}.
If $S$ is a saturation pulse with arbitrary Euler angles $\gamma(\omega)$ and  $\alpha(\omega)$,
none of the symmetry related pulses $S^\prime$ given in Table 
2 satisfies the Ramsey condition.
However, in principle
the simultaneous optimization of a pair of s$^2$-COOP Ramsey pulses 
that are not necessarily related by simple symmetry operations
(and that can also have different durations $T^{(1)}$ and $T^{(2)}$)
can result in individual pulses $S^{(1)}_{COOP}$ and $S^{(2)}_{COOP}$,
which satisfy neither the conditions for PP$_R$ nor for ST$_R$ pulses,
i.e., neither $\alpha_s(\omega)$ nor 
$\alpha^{nl}_a(\omega)$ are required to be zero for each of the individual pulses
(cf. Table 3).

\vskip 1em
(e) {\it Rectangular $90^\circ$ pulses}:
It is also of interest to include
in the following comparison simple
rectangular pulses, i.e.\ pulses with constant amplitude and phase
that are widely used in Ramsey-type experiments.
It is important to realize that simple rectangular pulses are only a
good approximation for UR(90$_y^\circ$) pulses for spins very close to resonance, i.e.\  with
offset frequencies  $\vert \nu \vert=\vert {\omega}/({2 \pi}) \vert \ll u^{max}$.
For larger offset frequencies up to about $\vert \nu \vert < u^{max}$,
simple
rectangular pulses
are still a
reasonable approximation of point-to-point pulses 
PP$_R$($z\to x$) pulses with 
\begin{equation}
\label{rrgamrectt}
R^{\it rect}_{\alpha}=2/\pi\approx 0.64,
\end{equation}
i.e.\ the linear part of the offset-dependent Euler angle $\alpha(\omega)$ is given by \cite{rectpulse_R}
\begin{equation}
\label{gamlinrect}
\alpha^{lin}(\omega)=  2 \omega T/\pi.
\end{equation}
Table 3 summarizes the
growing number of  degrees of freedom (indicated by black bullets) that are available in the pulse design process, when proceeding from broadband UR pulses via PP pulses and ST pulses to general s$^2$-COOP pulses.

\begin{table}
\begin{center}
\caption
{\label{tab:Freedom}{\bf Degrees of freedom for the Euler angles $\gamma$ and $\alpha$ characterizing the first pulse in different constructions  of Ramsey sequences.
}
}
\begin{tabular}{l l c c c c}
\\
  \hline
$S^{(1)}$ & $S^{(2)}$  & $\gamma(\omega)$ & $R_\alpha$ & $\alpha_s(\omega)$ &${\alpha}^{nl}_{a}(\omega) $ \\
  \hline
{\rm s}$^2$-COOP$^{(1)}$ & {\rm s}$^2$-COOP$^{(2)}$
  & $\bullet$ & $\bullet$ & $\bullet$ & $\bullet$  \\  
ST$_0$/ST$_R$
 & $S^{tr}$, $S^{tr}_{ps}$ & $\bullet$ & 0/$\bullet$ & $\bullet$ & 0  \\
PP$_0$/PP$_R$ %
 &$S^{tr}$,  $S^{tr}_{ps}$, $S^{tr}_{ip}$& $\bullet$ & 0/$\bullet$ & 0 & 0  \\
 UR &$S$, $S_{ps}$, $S_{ip}$, $S^{tr}$, $S^{tr}_{ps}$, $S^{tr}_{ip}$   & 0 & 0 & 0 & 0  \\
  \hline
 \hline
\end{tabular}
\begin{tabular}{l }
\footnotesize{
The symbol $\bullet$ indicates that the corresponding parameter is not restricted.}
\\
 \footnotesize{
The condition $\beta(\omega)=\pi/2$ applies to all Ramsey pulse types. \ \ \ \ \ \ \ 
\ \ \ \ \ \ \  \ \ \ \ \ \ \ \ \ \ \ 
}
\end{tabular}
\end{center}
\end{table}

\subsection{Comparison of performance as a function of pulse duration}

\begin{figure}[Hb!]
\centerline{
\includegraphics[width=.7\textwidth]{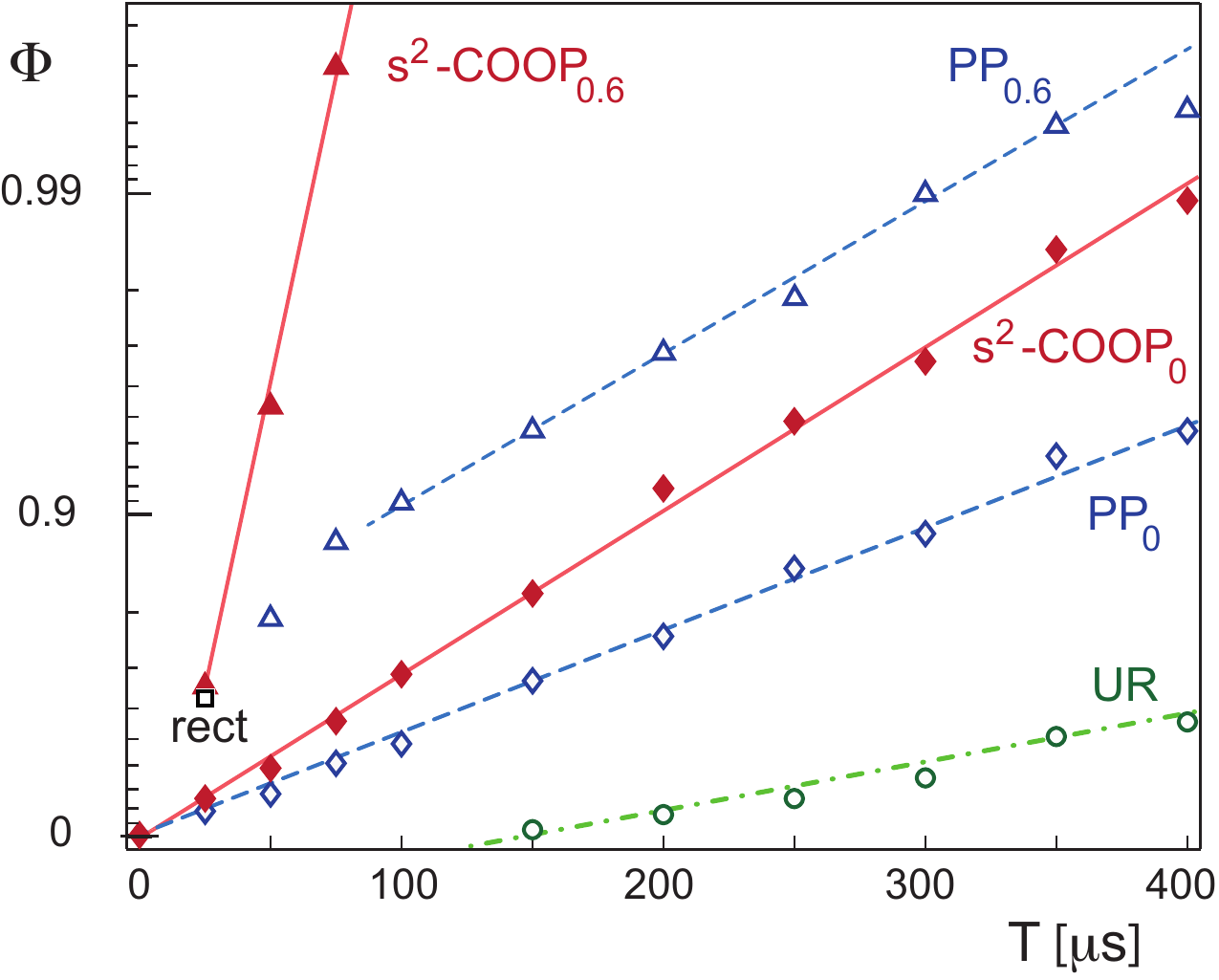} }
\caption{\small \label{fig:8} 
The figure shows the scaling of the performance $\Phi$ (cf.\ Eq.\ \eqref{defphi}) for different families of Ramsey sequences
as a function of pulse duration $T$ for a bandwidth that is 7 times larger than the maximum
pulse amplitude. 
} 
\end{figure}

The analysis presented in the previous sections
allowed us to clearly organize the
large number of possible Ramsey sequence constructions 
in terms of pulse type (UR, PP$_R$, ST$_R$, s$^2$-COOP$_R$)
and the symmetry relation between the two pulses.
From UR via PP$_R$ and ST$_R$ to s$^2$-COOP$_R$ pulses, the 
constraints for the individual pulses are more and more relaxed,
i.e.\ the number of available degrees of freedom
(represented by filled circles in Table 3)
increases.
The extent to which this translates in improved performance of Ramsey-type pulse sequences will be investigated in the following.
In fact, striking differences in performance are found for the different pulse sequence families.
The key results are summarized in Figure 8, which shows the achievable
broadband quality factor $\Phi$ (cf. Eq.\ \eqref{defphi})
as a function of pulse duration $T$. The extracted parameters of interest are summarized in Table 4.

In previous systematic studies of broadband UR \cite{UR_limits}, PP$_0$ \cite{BEBOP_limits} and 
PP$_R$ \cite{Iceberg} pulses,
it was empirically found that the performance 
$\Phi_P$ of a given pulse type
scales with pulse duration $T$ roughly as
\begin{equation}
\label{phiscale}
\Phi(T)
\approx 
1-c\  e^{-a T}
\end{equation}
with constants $a$ and $c$.
Hence, plotting 
$\tilde{\Phi}(T) =- {\rm ln}\{1-\Phi(T)\}$ as a function of $T$
is expected to approximately follow a straight line with slope $a$ and $y$-axis intercept $b=-{\rm ln}\{c\}$,
which is indeed the case (cf. Fig.\ 8 and Table 4).

Fig.\ 9 shows the desired 
Ramsey fringe pattern given by
Eq.\ \eqref{Mztarget}
(grey curve) for an effective evolution period of $\tau_{eff}=\tau+\delta$ of 95 $\mu$s  and the actual modulation of the $z$-component of the final
Bloch vector
$M^{\it final}_z(\nu)$
that can be achieved for a pulse duration of 75 $\mu$s
by different pulse families (Fig.\ 9 A, B, D, E, F) and 
by a rectangular $90^\circ$ pulse (Fig.\ 9 C) with the same maximum pulse amplitude of 10 kHz
and a corresponding pulse duration of  25 $\mu$s.

\begin{figure}[Hb!]
\centerline{
\includegraphics[width=.8\textwidth]{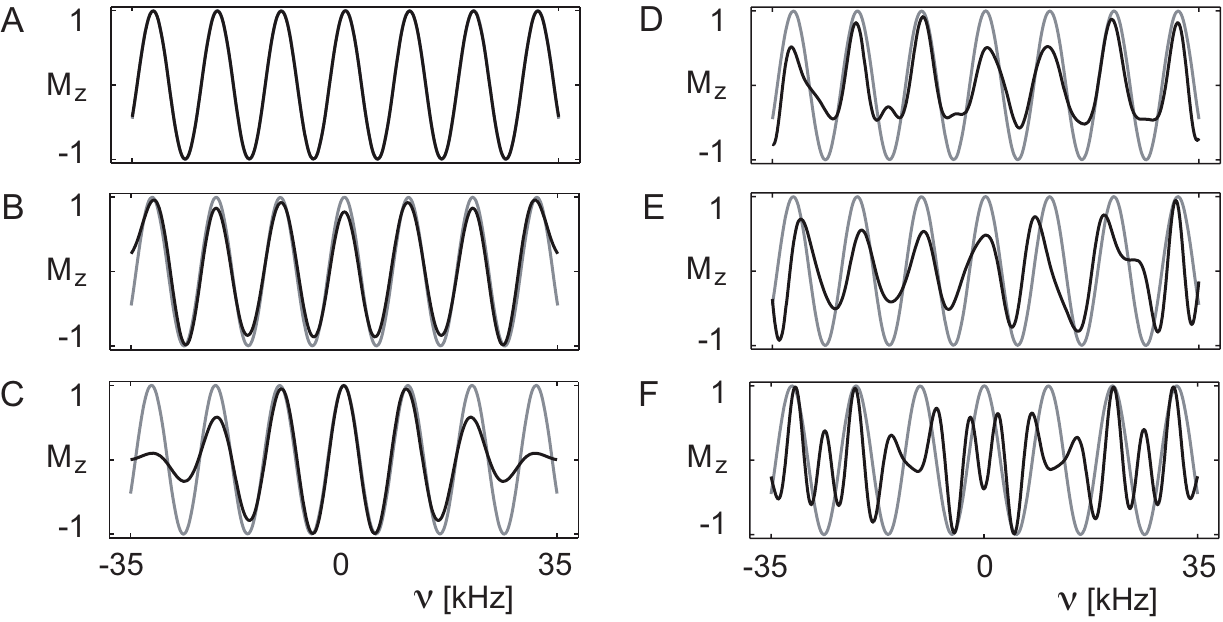} }
\caption{\small \label{fig:9} 
The grey and black curves show the desired ideal Ramsey  fringe pattern and the
simulated modulation of $M_z(\nu)$ for
s$^2$-COOP$_{0.6}$ (A) and  PP$_{0.6}$ pulses  (B),
for rectangular 90$^\circ$ pulses (C),
for
s$^2$-COOP$_0$ (D), 
PP$_0$ (E)
and UR pulses (F) (cf. Table 4) with a duration of
$T=75\ \mu$s (except for the rectangular pulses with $T=25\ \mu$s).
The effective evolution time $t_{\it eff}$ is 95 $\mu$s.
} 
\end{figure}

{\it Broadband UR pulses}:
For the desired range of offsets and scaling factors of the pulse amplitude, individual
UR(90$_y^\circ$) pulses were optimized using the
GRAPE algorithm as described in \cite{GRAPE, UR_limits}.
The maximum performance $\Phi(T)$ of $S$-$\tau$-$S^\prime$ Ramsey sequences, where $S$ is a UR($90^\circ_y$) pulse and $S^\prime=S^{tr}_{ip}$
is indicated in Fig.\ 8 by open circles.
Even for the longest considered pulse duration of $T$=400 $\mu$s, the quality factor is poor. 
As shown in Fig.\ 9 F,
for the best Ramsey constructions based on UR pulses with a duration of 75 $\mu$s,
the black curve representing $M^{\it final}_z(\nu)$  deviates strongly
from the desired
Ramsey fringe pattern (grey curve) over the entire offset range of interest.

 {\it Broadband PP$_0$ pulses}:
For the same pulse durations, a significantly improved performance
is found for $S$-$\tau$-$S^\prime$ Ramsey sequences, where $S$ is a PP$_0$($z\to x$) pulse 
 and $S^\prime=S^{tr}_{ip}$
as shown in Fig.\ 8 by open diamonds. 
The  PP$_0$ pulses were optimized using the
GRAPE algorithm described in \cite{GRAPE, BEBOP1}.
Note that  PP$_0$ pulses with a duration of 75 $\mu$s achieve better
quality factors than four times longer UR pulses.
However, as illustrated in Fig.\ 9 E, $M^{\it final}_z(\nu)$
still deviated significantly from the desired fringe pattern.
Fig.\ 10 D shows the detailed offset-dependent orientation of
the Bloch vectors after the first Ramsey pulse (before the SQF filter) and 
Figs.\ 10 D$^\prime$ and 10 D$^{\prime \prime}$ shows the Bloch vectors after the second Ramsey pulse
without and with ZQF filter, respectively.

 {\it Broadband s$^2$-COOP$_0$ pulses}:
 As shown by the filled diamonds in Fig.\ 8,
a significant further improvement of pulse sequence performance is
found for s$^2$-COOP$_0$ pulses 
and identical pulse durations $T^{(1)}=T^{(2)}=T$
that were optimized using the algorithm outlined in section 4.2.2.
In particular the 
slope 
$a$(s$^2$-COOP$_0)\approx 12$ ms$^{-1}$ %
is larger compared to $a$(PP$_0)\approx 7.3$ ms$^{-1}$ and
$a$(UR)$\approx 3.5$ ms$^{-1}$ (cf.\ Table 4).
Although for short durations the absolute gains are moderate,
this leads to markedly improved sequences for longer pulses that are required
to reach reasonabaly good quality factors.
\begin{figure}[Hb!]
\centerline{
\includegraphics[width=1.0\textwidth]{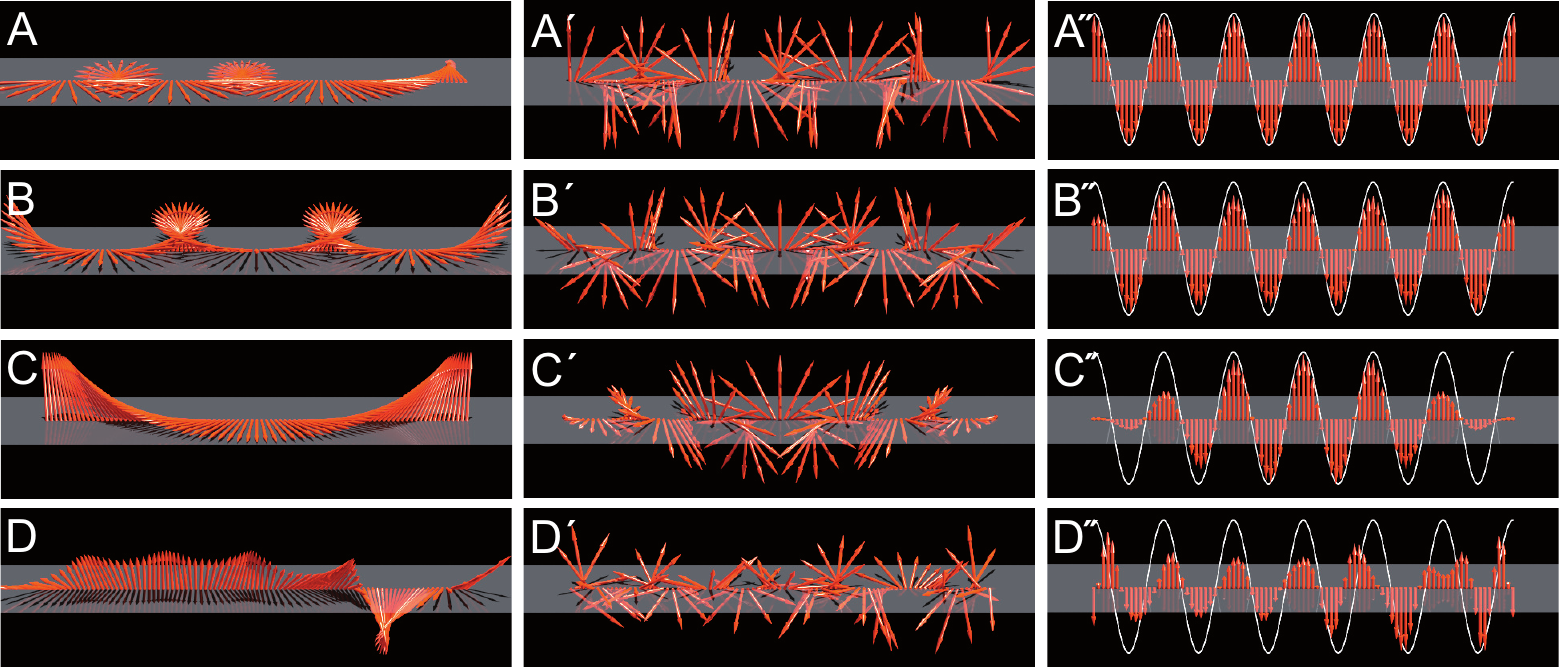} }
\caption{\small \label{fig:10} 
Panels A-D show the
orientation of
the Bloch vectors after the first Ramsey pulse before the SQF filter is applied
for the pulses shown in Fig.\ 6 A-D, repectively:
(A) s$^2$-COOP$_R$ pulses with $R=0.53$, (B) PP$_R$  pulses with $R=0.53$,
(C) rectangular 90$^\circ$ pulses  and 
(D) 
PP$_0$ pulses. 
The range of offsets is $-35$ kHz $\leq \nu \leq$ 35 kHz.
The corresponding  Bloch vectors after the second Ramsey pulse
without (A$^\prime$-D$^\prime$) and with (A$^{\prime \prime}$-D$^{\prime \prime}$)
ZQF filter are shown in the middle and right panels, respectively. The effective evolution time $t_{\it eff}$ is 85 $\mu$s and the
dashed curves in the right panels represents the target Ramsey fringe pattern.
} 
\end{figure}
As the relative slope $R$ of s$^2$-COOP$_0$ pulses is zero by definition, the auxiliary delay $\delta$ is also zero and the effective
evolution time $t_{\it eff}$ of the Ramsey sequence
is identical to the inter-pulse delay $\tau$ (as for UR and PP$_0$ pulses).
The shapes of the pulse components $u_x(t)$ and $u_y(t)$
for  s$^2$-COOP$_0$ pulses 
with a duration of 75 $\mu$s
are displayed in Figs.\ 6  D.

 {\it Rectangular pulses}:
At first sight, it may be surprising that the quality factor
$\Phi=0.62$
 of a  Ramsey sequence based on simple rectangular pulses (indicated by an open square in Fig.\ 8) is markedly better than
the performance based on highly optimized s$^2$-COOP$_0$ pulses of comparable duration.
However, this is a simple result of the
fact that in contrast to UR, PP$_0$ and s$^2$-COOP$_0$ pulses, which yield
Ramsey sequences with a vanishing auxiliary delay ($\delta=0$),
$\delta$ is  {\it not} zero for rectangular 90$^\circ$ pulses.
This ensues from the non-vanishing linear part of $\alpha(\omega)$ of rectangular
90$^\circ$ pulses \cite{rectpulse_R} with a relative slope of   
$R^{\it rect}_\alpha\approx 0.64$, cf.\ Eq.\ \eqref{rrgamrectt}.
This is illustrated in Fig.\ 6 C, where the effective evolution time
$t_{\it eff}=85$ $\mu$s consists of an inter-pulse delay $\tau=53$ $\mu$s
and the effective auxiliary delay $\delta= 2 \cdot R^{\it rect}_\alpha\cdot 25$ $\mu$s=32 $\mu$s. 
The offset-dependent positions of the Bloch vectors after the first rectangular pulse are shown in Fig.\ 10 C. Note that for offsets
$\vert \nu \vert > u^{max}$=10 kHz, the Bloch vectors still have significant $z$-components (that will be eliminated by the following SQF filter) because
the Euler angle $\beta(\omega)$ does deviate considerably from the desired value of $90^\circ$ for these offsets. 
Figs.\ 10 C$^\prime$ and C$^{\prime \prime}$ show the orientation of the Bloch vectors after the 
second rectangular pulse 
without and with zero-quantum filter, respectively. 
For $t_{\it eff}=95$ $\mu$s, 
the resulting modulation of its $z$-component is also represented by the black curve in Fig.\ 9 C. As expected, the desired ideal fringe pattern indicated by the grey curve is closely matched for small offsets $\vert \nu \vert$, but cannot be approached if $\vert \nu \vert$ is larger than $u^{max}$.
As pointed out in section 2.1, for many applications the non-vanishing
auxiliary constant delay $\delta$ of rectangular pulses is acceptable
and it is particularly interesting to see the impact on the performance of Ramsey sequences
if the condition $R=0$
(and hence $\delta=0$) is lifted for PP$_{R}$, ST$_{R}$ and
s$^2$-COOP$_R$ pulses.

 \vskip 1em
 {\it Broadband PP$_R$ pulses}:
  The open circles in Fig.\ 5 show the Ramsey quality factor $\Phi$
of PP$_R$ pulses $S$ (and  $S^\prime=S^{tr}_{ip}$)
of duration $T=75$ $\mu$s that were optimized for different slopes $R$ based on the GRAPE-based approach described in \cite{Iceberg}.
Their performance is significantly better than the quality factor $\Phi$ 
that is reached by rectangular pulses, which is  marked in Fig.\ 5 by an open rectangle.
The best performance of the PP$_R$ pulses  ($\Phi=0.88$) is found
for $R=0.6$. The pulse shape for $R=0.53$  (with $\Phi=0.87$) is shown in Fig.\ 6 B.
The pulse has a vanishing $x$-component ($u_x(t)=0$) and the
$y$-component  $u_y(t)$ alternates between the values $\pm u^{\it max}$.
The effective evolution time
$t_{\it eff}=85$ $\mu$s consists of an inter-pulse delay $\tau=5.5$ $\mu$s
and the effective evolution time during the pulses is $\delta= 2 R \cdot 75\mu$s=79.5$\mu$s. 
The offset-dependent orientation of the Bloch vectors after the first and the second pulse are shown in Fig.\ 10 B and B$^\prime$ (and in B$^{\prime \prime}$ after the ZQF).
Figs.\ 9 B and 10 B$^{\prime \prime}$ demonstrate that 
the desired fringe pattern is approached 
for the entire offset range of interest. Compared to
the case of rectangular pulses, the overall deviation from the ideal pattern are smaller and evenly distributed over the entire offset range,
because the gradients for all offsets were given the same weight in the optimization of the PP$_R$ pulses.
As the best value of $R$ for PP$_R$ pulses with a duration of $T=75$ $\mu$s
was 0.6, 
PP$_{0.6}$ pulses were also optimized for other pulse durations $T$ and the resulting
Ramsey sequence quality factors $\Phi$
are shown by open triangles in Fig.\ 8.

\begin{table}
\begin{center}
\caption
{\label{tab:scalingfit}{\bf
Comparison of the auxiliary delay $\delta$ and scaling parameters of quality factor $\Phi(T)$ for the Ramsey pulse families shown in Fig.\ 8.
}
}
\begin{tabular}{l l c c c c l c}
\\
  \hline
$S^{(1)}$ & $S^{(2)}$  & $\delta$ [$\mu$s] & $a$ [ms$^{-1}$] & $b$ & c& $\Phi(T=75$ $\mu$s)
& $T_{\Phi=0.996}$ [$\mu$s] \\
  \hline
  $S^{(1)}_{\rm COOP_{0.6}}$ & $S^{(2)}_{\rm COOP_{0.6}}$  
  &90   &   90  & -1.3 & 3.7&\ \ \ \   0.996 & 75
   \\  
ST$_{0.6}$ & $S^{tr}_{ps}$
  & 90  &   90  & -1.3 & 3.7& \ \ \ \  0.996 & 75
\\
PP$_{0.6}$ & $S^{ip}_{ip}$
  &  90  &   11  & 1.3 &0.27
 & \ \ \ \  0.88 & $\approx$400
\\
  $S^{(1)}_{\rm COOP_{0}}$ & $S^{(2)}_{\rm COOP_{0}}$  
  &  0  &   12  & 0 &1 & \ \ \ \  0.56  & $\approx$460
   \\  
   ST$_{0}$ & $S^{tr}_{ps}$
  &  0  &   12  & 0 &1 &\ \ \ \   0.56  &$\approx$460
\\
  PP$_{0}$ & $S^{ip}_{ip}$
  &  0  &   7.3  & 0 & 1&\ \ \ \   0.4 & $\approx$750
\\ 
   UR & $S^{ip}_{ip}$
  &  0  &   3.5  & -0.5 & 1.65 &\ \ \ \   $< 0$ & $\approx$1700
\\   
     \hline
     \hline
\end{tabular}
\begin{tabular}{l}
\footnotesize{
The auxiliary delay $\delta$ corresponds to a pulse duration $T=75$ $\mu$s. For each pulse type, the slope $a$  } \\
\footnotesize{
and axis intercept $b$ are determined from Fig. 8
 and  $c=e^{-b}$. With these parameters, the quality}\\
 \footnotesize{
factor scales with $T$ approximately as
 $\Phi\approx 1-  e^{-a T-b}= 1-c\  e^{-a T}$.
}
\end{tabular}
\end{center}
\end{table}

 {\it Broadband ST$_R$ and s$^2$-COOP$_R$ pulses}
As discussed in 
section 5  for the pulse duration $T=75$ $\mu$s
the 
best quality factor of $\Phi=0.9965$
is achieved for $R=0.53$. 
However, as rectangular pulses and the best
PP$_{R}$ pulses have values of 
$R\approx0.6$,
we also chose this value (corresponding to $\Phi=0.9960$)
for the comparison of the performance as a function of pulse duration $T$ in Fig.\ 8.
This figure illustrates the far superior
performance that can be achieved by 
s$^2$-COOP$_R$ pulses compared to rectangular pulses,
PP$_R$ pulse and UR pulses.

Fig.\ 10 A shows that the first pulse brings the Bloch vectors almost
 completely into the transverse plane for all offset frequencies. This figure also
 illustrates the
 nonlinear phase roll, which provides significantly more flexibility in the pulse optimization. Although the offset-dependent orientations of the 
 Bloch vectors after the second pulse appear to be rather chaotic (cf.\ Fig.\ 10 A$^\prime$), their $z$-components
do  approach the desired Ransey fringe pattern
 with outstanding fidelity (cf.\ Fig.\ 10 A$^{\prime \prime}$). In fact,  in Fig.\ 9 A, $M^{\it final}_z(\nu)$ and
 the ideal Ramsey fringe pattern are indistinguishable due to the excellent match.

 The s$^2$-COOP$_R$ pulse shapes for $R=0.53$ are displayed in Fig.\ 6 A.
 As discussed in sections 5 and 6.2,
the s$^2$-COOP$_R$ pulse pair
corresponds to a very good approximation to pairs of
 ST$_R$ pulses with $S^\prime=S^{\it tr}_{\it ps}$.
 In fact, for the optimization parameters considered here,
the performance of s$^2$-COOP$_R$ Ramsey pulses
is closely approached by the family of ST$_R$ pulses.

Fig.\ 8 demonstrates that the excellent quality factor $\Phi$
of s$^2$-COOP$_R$ and ST$_R$
pulses
not only exceed by far the performance of simple rectangular pulses
but also results in ultra short pulses compared to conventional approaches
based on individually optimized pulses.
The line fitting the data in Fig.\ 8 corresponds to an extremely 
steep slope of 
$a$(s$^2$-COOP$_{0.6})\approx 90$ ms$^{-1}$ (and $y$-axis intercept $b$(s$^2$-COOP$_{0.6})\approx -1.3$).
Note that a Ramsey quality factor $\Phi>0.996$ can be achieved by
s$^2$-COOP$_R$ and ST$_R$ pulses with a duration $T=75$ $\mu$s,
which is only {\it three} times longer than the duration of a rectangular 90$^\circ$ pulse.
For comparison, based on a simple extrapolation of 
the data shown in Fig.\ 8,
a comparable quality factor 
is expected to require durations in the order of
$T\approx400$ $\mu$s for PP$_R$ pulses,
$T\approx750$ $\mu$s for PP$_0$ pulses,
and about 1.7 ms for UR pulses,
corresponding to %
16, 
30 and more than 70 times the duration 
of a rectangular 90$^\circ$ pulse, respectively.

\section{Experimental demonstration}

As pointed out in section 1.1,
the Ramsey sequence plays an important role in many fields,
including 2D NMR spectroscopy, where it is used as a standard frequency-labeling
building block in many experiments \cite{Ernst}.
The specific parameters (maximum pulse amplitude, desired bandwidth of frequency offsets, etc.) of the optimization problem defined in section 5
were motivated by applications of 2D nuclear Overhauser enhancement spectroscopy (NOESY), where the bandwidth of interest is much larger than the 
maximum available pulse amplitude.
More specifically, it was assumed that the desired bandwidth is seven times larger than the maximum control amplitude of 10 kHz, corresponding e.g. to 
 $^{13}$C-$^{13}$C-NOESY experiments at high magnetic fields.

 \begin{figure}[Hb!]
\centerline{
\includegraphics[width=.55\textwidth]{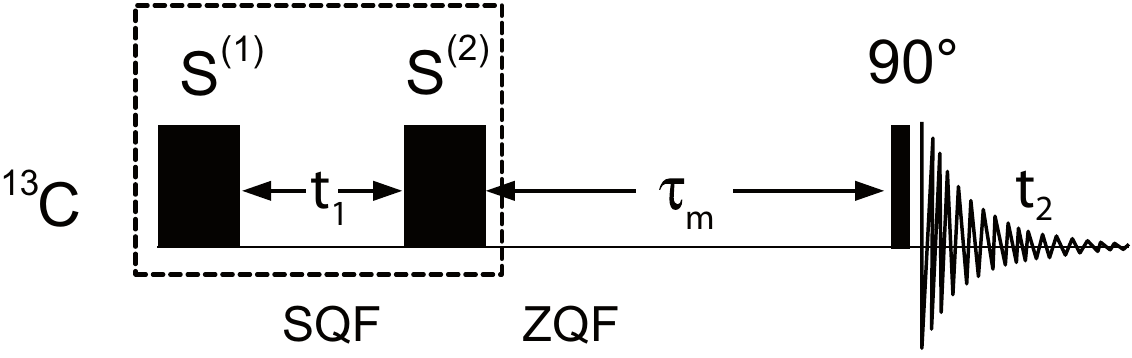} }
\caption{\small \label{fig:11} 
Schematic representation of the 
2D-$^{13}$C-$^{13}$C-NOESY  pulse sequence \cite{CC_NOESY_2} that
was used to demonstrate the performance of different types of Ramsey pulses
$S^{(1)}$ and $S^{(2)}$ 
 which form the frequency labeling element of the NOESY experiment  indicated by the dashed rectangle.
$^1$H spins were decoupled using 
the composite-pulse sequence WALTZ-16 \cite{Waltz-16} (not shown).
} 
\end{figure}
 
To test the outstanding theoretical properties of 
s$^2$-COOP sequences in practice, we performed
2D-$^{13}$C-$^{13}$C-NOESY experiments on a Bruker AV III 600 spectrometer 
with a magnetic field strength of 14 Tesla using a sample of $^{13}$C-labeled $\gamma$-D-glucose dissolved in Dimethylsulfoxid (DMSO).
The $^{13}$C-$^{13}$C-NOESY pulse sequence from \cite{CC_NOESY_2} was used (without the $^{15}$N-decoupling pulses which were not necessary
for the glucose test sample). 
The zero-quantum filter (ZQF) after the frequency labeling building block, i.e. after the second Ramsey pulse (cf. Fig.\ 11) was implemented
by a standard chirp pulse/gradient pair \cite{ZQF_Keeler}.
The complete pulse sequence of the $^{13}$C-$^{13}$C-NOESY experiment is shown schematically in Fig.\ 11, where the $S^{(1)}$-$\tau$-$S^{(2)}$
Ramsey-type frequency-labeling building block 
is indicated by the dashed box and the inter-pulse delay $\tau$ corresponds to the
evolution period that is usually called "$t_1$" in 2D NMR.

In the experiments, 
the two 90$^\circ$ pulses of the Ramsey building block  were implemented 
by the following three pulse sequences:
the s$^2$-COOP$_{0.6}$ pulse pair with a duration
$T=3/(4 u^{max})$ (which is three times longer the duration of a rectangular 90$^\circ$ pulse
with the same maximum pulse amplitude $u^{max}$), 
the PP$_{0.6}$ pulse pair  with
the same pulse duration
and 
as a pair of standard rectangular pulses.

The chemical shift range of 40 ppm for the $^{13}$C glucose sample at 14 Tesla
corresponds to a bandwidth of about 6 kHz.
As the Ramsey pulses were optimized for the challenging case of $\Delta \nu=7 u^{max}$,
a correspondingly scaled maximum pulse amplitude of $u^{max}=0.86$ kHz
was used in the demonstration experiments.
For this amplitude,
the pulse durations $T$ were 870 $\mu$s
for the
s$^2$-COOP$_{0.6}$ and PP$_{0.6}$ pulses 
and
290 $\mu$s for the rectangular pulses.
In all experiments, the final detection pulse after the NOESY mixing period
$\tau_{mix}$ was a strong rectangular 90$^\circ$ pulse
with a pulse amplitude of 12.2 kHz (and a corresponding duration of 20.5 $\mu$s), which was sufficient to cover the bandwidth of 6 kHz.
For larger bandwidths, this pulse could be replaced by an optimized broadband excitation pulse \cite{BEBOP_limits}. 

The NOESY mixing time was $\tau_{mix}=50$ ms and the recycle delay 
between scans was 280 ms. The spectra were recorded at a temperature of 293 K, using a TXI probe with 512 $t_1$ increments, 16 scans for each increment
and 8k data points in the detection period $t_2$.
As the minimum 
$t_1$ value is given by $\delta=2 R T=2 \cdot 0.6\cdot 870$ $\mu$s$\ =\ $1.04 ms
(cf. Eq.\ \eqref{d2rt}),
the  time-domain data was completed
using standard backward linear prediction \cite{back_lin_pred}  
before the
2D Fourier transform.
The processing parameters for all spectra were identical. Selected slices of the $^{13}$C-$^{13}$C NOESY spectra for 
s$^2$-COOP$_{0.6}$, PP$_{0.6}$ and conventional rectangular
pulses are displayed in Fig.\ 12 A-C.   %

 \begin{figure}[Hb!]
\centerline{
\includegraphics[width=.7\textwidth]{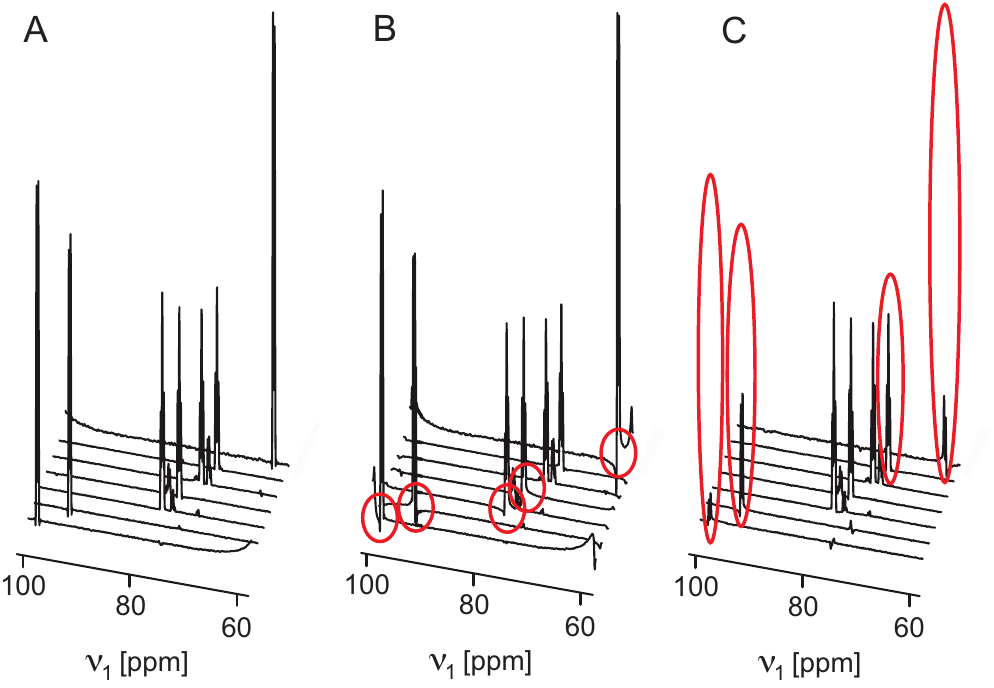} }
\caption{\small \label{fig:12} 
The figure shows cross sections of 2D-$^{13}$C-$^{13}$C-NOESY experiments
of  $^{13}$C-labeled $\gamma$-D-glucose. 
The pulses $S^{(1)}$ and $S^{(2)}$ of the Ramsey-type frequency-labeling block indicated 
by the dashed box in Fig.\ 11
were 
s$^2$-COOP$_{0.6}$ pulses (A),  PP$_{0.6}$  pulses (B)
and   rectangular 90$^\circ$ pulses (C).
In B, the ellipses indicate signal distorsions due to
phase errors. In C, the ellipses point out amplitude losses relative to A.} 
\end{figure}

 \begin{figure}[Hb!]
\centerline{
\includegraphics[width=.7\textwidth]{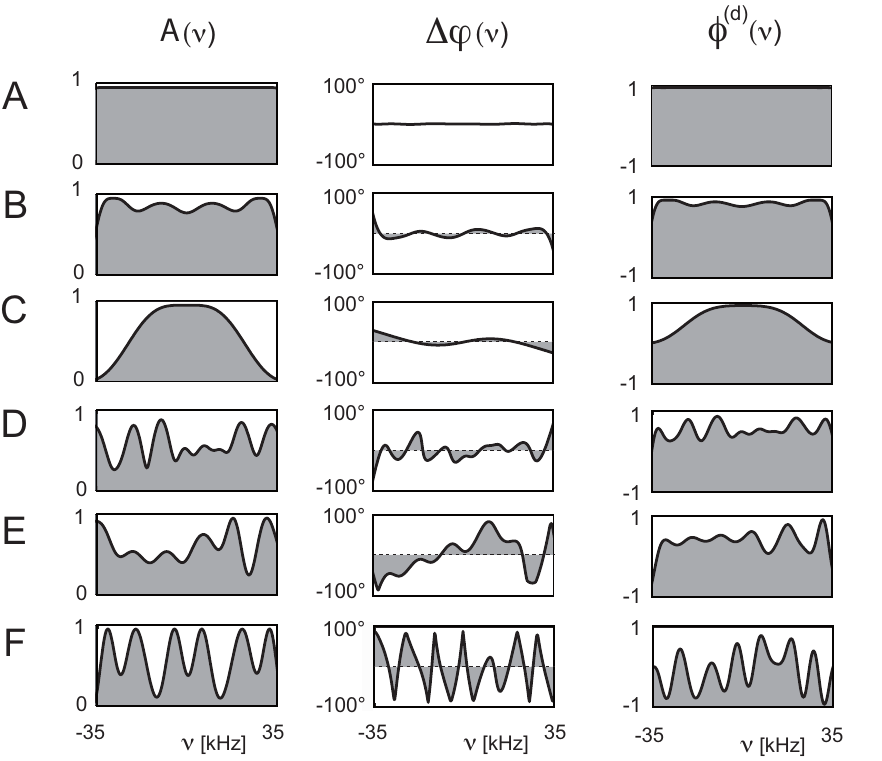} }
\caption{\small \label{fig:13} 
The simulated  signal amplitude 
$A(\nu)$ (left column)
and phase error $\Delta \varphi(\nu)$ in the $\nu_1$ dimension (middle column)
of a 2D-NOESY spectrum is shown as a function of offset frequency $\nu$
for
s$^2$-COOP$_{0.6}$ pulses (A),  PP$_{0.6}$  pulses (B),
rectangular 90$^\circ$ pulses (C),
s$^2$-COOP$_{0}$ pulses (D), PP$_{0}$  pulses (E)
and UR pulses (F).
The right column shows the offset-dependent Ramsey quality factor
$\phi^{(d)}(\nu)$, %
which reflects both pulse amplitude
and phase errors.
} 
\end{figure}

As expected from the simulations shown in Fig.\ 9 C,
rectangular pulses perform well
for relatively small offsets frequencies (corresponding to the center of the spectra in Fig.\ 
12, i.e. to the signals in the chemical shift range from 70 to 80 ppm).
However,
for large offsets from the irradiation frequency at the center of the spectrum,
the performance of rectangular pulses breaks down (cf. Fig.\ 9 C),
resulting in a dramatic signal loss for the peaks in the NOESY experiments that are located at the edge of the spectral range (near 60 ppm and 100 ppm) in Fig.\ 12 C.
In contrast, s$^2$-COOP$_{0.6}$ pulses also perform perfectly well for large offsets (cf. Fig.\ 9 C), resulting in large gains of up to an order of magnitude
for the signal amplitudes at the edge of the spectral range.
As expected from the simulated fringe patterns in Fig.\ 9 B,
PP$_{0.6}$ pulses also 
yield significantly increased signal amplitudes for large offset frequencies compared to rectangular pulses. 
However, as shown in Fig.\ 12 B, the peaks also have relatively large phase errors, resulting in asymmetric line-shapes and baseline distortions close to large peaks (indicated by the ellipses).
The corresponding simulated
offset dependence of the
signal amplitude $A(\nu)$
and of the
phase error $\Delta \varphi (\nu)$ in the $\nu_1$ dimension
(corresponding to the evolution period $t_1$ in the time domain) can be calculated based on the Euler angles
$\beta^{(1)}$ and $\beta^{(2)}$ and the nonlinear components of 
$\alpha^{(1)}$ and $\gamma^{(2)}$ as
\begin{equation}
\label{ampldef}
A(\nu)= \sin\{\beta^{(1)}(\nu)\} \sin\{\beta^{(2)}(\nu)\}
\ \ \ {\rm and} \ \ \ 
\Delta \varphi (\nu)=\alpha^{(1)nl} (\nu)
+\gamma^{(2)nl} (\nu)
\end{equation}
(cf.\ Eq.\ \eqref{g1nlg2nl}) 
and are shown in the left and middle  panels of Fig.\ 13 A-F, respectively. 
The quality factor 
$\Phi^{(d)}(\nu)$ defined in Eq.\ \eqref{Phidw} can also be expressed 
in terms of
$A(\nu)$
and  $\Delta \varphi (\nu)$
as
\begin{equation}
\label{delphidefx}
\Phi^{(d)}(\nu)=A(\nu) \cos\{\Delta \varphi (\nu)\},
\end{equation}
i.e.\ it 
reflects both $A(\nu)$ and $\Delta \varphi (\nu)$
as shown in panels of Fig.\ 13 A-F.
We note in passing that for applications with specific weights $w_A$ and $w_{\Delta \varphi}$ for amplitude and phase errors,
a tailor-made quality factor
\begin{equation}
\label{phie}
\Phi^{e}(\nu)=
1-
w_A
\{1-A(\nu)\}^2
-
w_{\Delta \varphi}
\{\Delta \varphi (\nu)\}^2
\end{equation}
could be used in the optimizations.

 \begin{figure}[Hb!]
\centerline{
\includegraphics[width=.6\textwidth]{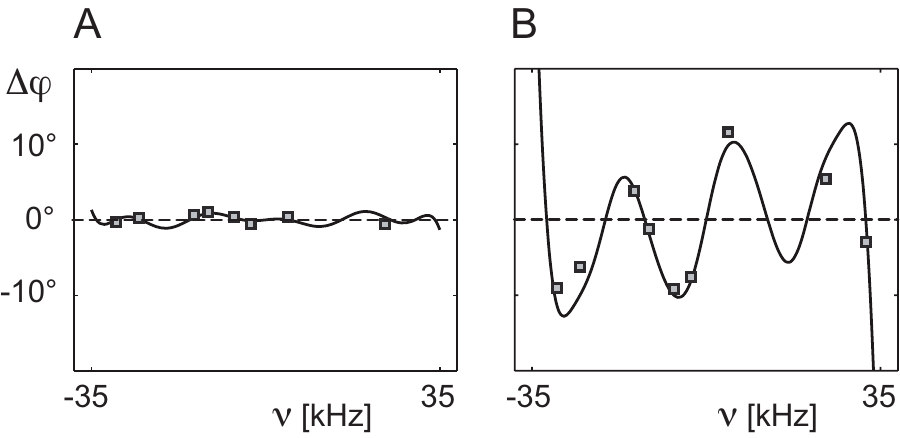} }
\caption{\small \label{fig:14} 
Expanded view of the simulated phase errors shown in Fig.\ 13 A and B for 
s$^2$-COOP$_{0.6}$ pulses  and  PP$_{0.6}$  pulses, respectively (black curves).
Experimentally measured phase errors $\Delta \varphi(\nu)$
based on the spectra displayed in
Figs. 12 A and B
are shown by open squares.} 
\end{figure}

Whereas
the s$^2$-COOP$_{0.6}$ pulses create almost ideal signal amplitudes $A(\nu)\approx 1$ (cf. left panel of Fig.\ 13 A)
 for
all frequencies $\nu$ in the optimized offset range, for the PP$_{0.6}$ pulses
the signal amplitude varies between 0.7 and 0.9  %
 (cf.\ left panel of Fig.\ 13 B).
Similarly,
the phase errors $\Delta \phi(\nu)$ are smaller than 1.5$^\circ$ for the
s$^2$-COOP$_{0.6}$ pulses 
(cf. middle panel of Fig.\ 13 A), whereas noticeable phase errors of more than $\pm 10^\circ$ are created by the 
PP$_{0.6}$ pulses (cf. middle panel of Fig.\ 13 B).
Panels A and B in Fig.\ 14 show 
enlarged views of these phase errors. In addition to the
simulated curves, Fig.\ 14 also 
shows 
experimentally determined phase errors (open squares) based on the spectra displayed in
Figs.\ 12 A and B.
A reasonable match is found between experimental and simulated data,
confirming the superior performance of s$^2$-COOP$_{R}$
pulses compared to conventional PP$_{R}$ and rectangular pulses
in broadband Ramsey-type pulse sequences.
The pulses with $\delta=0$ (corresponding to $R=0$) have significantly
poorer performance both in terms of signal amplitude and phase
for the same pulse durations, as shown in Fig.\ 13
 D-F.

\section{Conclusions and outlook}

Here, we introduced the concept of s$^2$-COOP pulses that are optimized simultaneously and act in a cooperative way in the same scan. Pulse cooperativity within the same scan (cf.\ Fig.\ 1 A)
complements the multi-scan COOP approach introduced in \cite{COOP_1} (cf. Fig.\ 1 B).
A general filter-based approach was introduced in section 4.1 that makes it possible to simultaneously optimize
an {\it arbitrary} number of s$^2$-COOP pulses. This makes it possible to optimize entire pulse sequences, rather than isolated pulses. The proposed s$^2$-COOP quality factors are based on the desired transfer function of the pulse sequence, which is essentially a {\it product} of the 
transfer functions of the individual pulses and filter elements. 
This is in contrast to the tracking approach for the optimization of 
decoupling sequences \cite{dec_Jorge, dec_Angew, dec_scaling}, where the overall
performance of a multiple-pulse sequence depends on the {\it sum} of the deviations from the ideal transfer function
during the pulse sequence. 

As an illustrative example of s$^2$-COOP pulses, we analyzed the important class of Ramsey-type experiments. Based on this analysis, a symmetry-adapted approach for the 
optimization of s$^2$-COOP pulse {\it pairs} for Ramsey sequences was discussed in section 4.2 that provides a different perspective and additional insight into this optimization problem. However, it is limited to the optimization of {\it two} pulses, in contrast to the general filter-based approach discussed in  section 4.1, which does not have this limitation.
The development of s$^2$-COOP Ramsey sequences provides excellent ultra short broadband pulses with a bandwidth that can be much
larger than the maximum available pulse amplitude.
In the chosen example, the bandwidth was seven 
times larger than the pulse amplitude, but the 
proposed algorithms can of course also be applied to even larger bandwidths.
Compared to conventional approaches based on the isolated optimization of individual pulses such as
universal rotation (UR) pulses \cite{UR_limits},  point-to point pulses with constant phase of the final magnetization as a function of offset (called PP$_0$ pulses)
and pulses that create a linear phase slope as a function of offset (called PP$_R$ pulses or Iceberg pulses \cite{Iceberg}),
the minimum pulse duration to reach the required overall performance of a Ramsey experiment
is up to two orders of magnitude shorter for s$^2$-COOP.
Decreased pulse durations result in reduced relaxation losses during the pulses, less experimental imperfections and also less sample heating, which is  particularly important for 
{\it in vivo} spectroscopy and applications in medical imaging.
The analysis of the resulting Ramsey s$^2$-COOP pulses also lead to the discovery of the powerful class of ST$_R$ pulses discussed in section 6, which makes it possible to construct Ramsey sequences based on the individual optimization of pulses, closely approaching the performance of s$^2$-COOP pulses for the
optimization parameters considered here.

 \begin{figure}[Hb!]
\centerline{
\includegraphics[width=.6\textwidth]{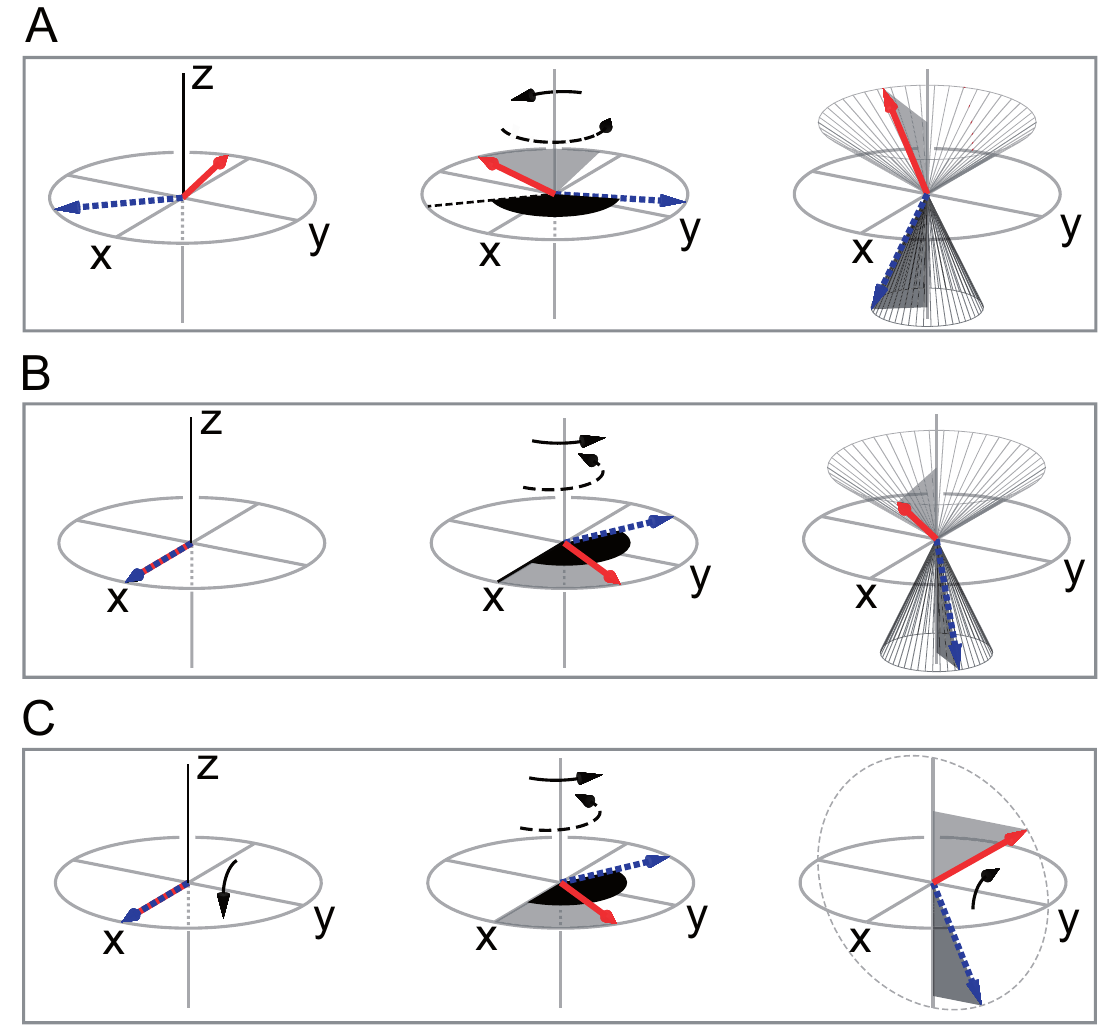} }
\caption{\small \label{fig:15} 
Graphical representation of the orientation of
 two exemplary Bloch vectors with
  different offset frequencies $\omega$ 
after the first Ramsey pulse (left), after the inter-pulse delay $\tau$ (middle)
and after the second Ramsey pulse (right) for the same effective evolution time $t_{\it eff}$.
Panel A corresponds to 
the case of s$^2$-COOP$_{R}$ 
and PP$_{R}$ pulses with $R \ne 0$
and to s$^2$-COOP$_{0}$, where the vectors are not necessarily oriented
along the $x$-axis after the first pulse.
Panel B corresponds to 
the case of
PP$_{0}$ pulses 
and panel C corresponds to 
the case of
UR pulses and ideal hard pulses. 
} 
\end{figure}

When comparing Ramsey pulses with the {\it same} duration $T$, 
the significant performance gain from UR via PP$_R$ to ST$_R$ and s$^2$-COOP$_R$ pulses demonstrated in Figs. 8, 9 and 13 
is strongly correlated with the increasing number of degrees of freedom (cf.\ Table 3) for the 
offset-dependent Euler angles of these pulse types.

 \begin{figure}[Hb!]
\centerline{
\includegraphics[width=.5\textwidth]{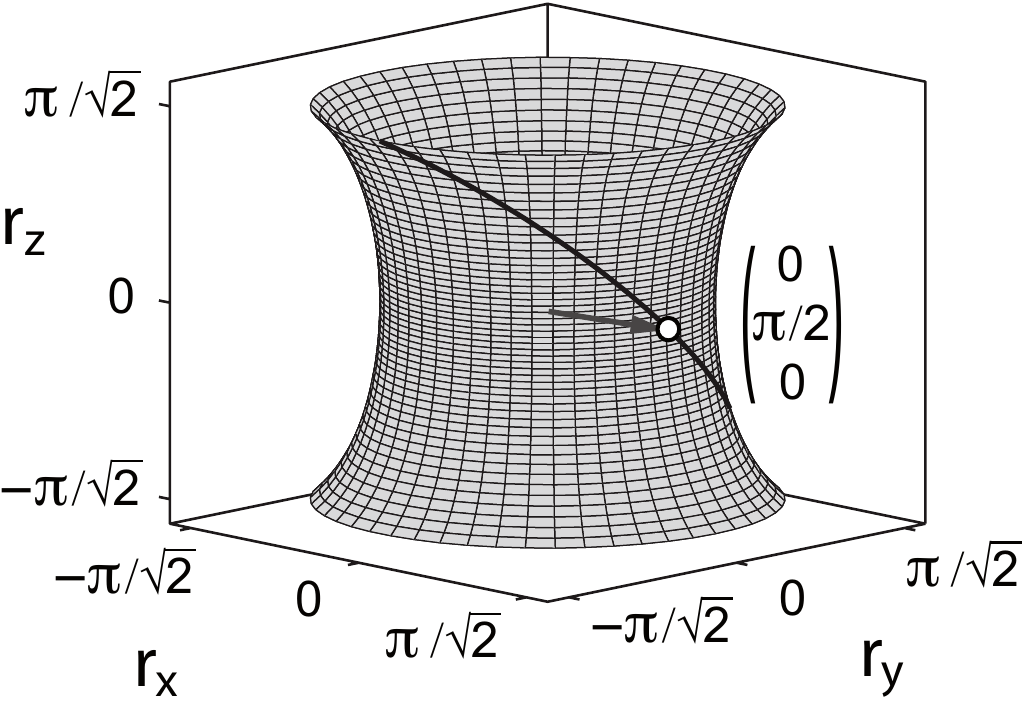} }
\caption{\small \label{fig:16} 
The figure shows the allowed effective rotation vectors ${\bf r}$ for different types of Ramsey pulses $S^{1}$,
where the length of ${\bf r}$ is given by the rotation angle (in units of radians) and its orientation corresponds to the rotation axis. %
The arrow 
represents a rotation by $90^\circ$ around the $y$-axis, corresponding to UR pulses and ideal
hard pulses. 
In the case of PP$_0$ pulses, the Bloch vectors are also brought from $z$ to $x$ (Fig.\ 15 B), but the rotation axis is {\it not} fixed to the $y$-axis \cite{UR_limits, URconstruction}.
The black curve illustrates the locations of the tips of the allowed effective rotation vectors
of PP$_0$ pulses.
For 
 PP$_{R}$,  ST$_{R}$  and
s$^2$-COOP$_{R}$ pulses, 
the effective rotation vectors of the individual Ramsey pulses can be located anywhere on the grey surface %
\cite{URconstruction, Kuehlturm}.
} 
\end{figure}

It is also instructive to consider the increasing flexibility in terms of the effective rotation vectors of the individual pulses and the trajectories of Bloch vectors during the Ramsey sequence.
Fig.\ 15 schematically 
displays the orientations of two exemplary Bloch vectors with different offset frequencies $\omega$ 
after the first Ramsey pulse (left), after the inter-pulse delay $\tau$ (middle)
and after the second Ramsey pulse (right).
For ideal UR pulses, which are most restrictive, the effect of the first pulse is a $90^\circ$ rotation around the $y$-axis, bringing both initial Bloch vectors from the $z$-axis to the $x$-axis (Fig.\ 15 C). The corresponding rotation vector ${\bf r}$ of the UR($90^\circ_y$)  pulse 
with components $r_x=0$, $r_y=\pi/2$ and $r_z=0$ 
is represented in Fig.\ 16 by an arrow
and the location of its tip is indicated by a circle.
 In the case of PP$_0$ pulses, the Bloch vectors are also brought from $z$ to $x$ (Fig.\ 15 B), but the rotation axis is {\it not} fixed to the $y$-axis \cite{URconstruction, UR_limits}.
For example, a rotation by 180$^\circ$ around the axis ${1\over{\sqrt{2}}}({\bf e}_x+{\bf e}_z)$ (corresponding to the bisecting line of the angle
between the $x$ and $z$-axis) has the same result
and the black curve in Fig.\ 16 
indicates the set of all rotation vectors 
that are compatible with a PP$_0$($z\rightarrow x$) pulse.
In the case of PP$_R$, ST$_R$ and s$^2$-COOP$_R$ pulses,
the first pulse is allowed to bring the Bloch vectors from the north pole (i.e. from the $z$-axis) to different locations on the equator of the Bloch sphere (cf. Fig.\ 15 A).
Hence, the allowed rotation vectors are not limited to the black curve in Fig.\ 16, but can be located anywhere on the 
grey surface \cite{URconstruction, Kuehlturm}. Whereas for PP$_R$ pulses the angles between the $x$-axis and the Bloch vector on the equator (i.e. the phase of the Bloch vector, which is identical to the Euler angle $\alpha(\omega)$) 
is required to be a linear function of the offset frequency (cf. Table 3), this condition is lifted for ST$_R$ and s$^2$-COOP$_R$ pulses.
As illustrated in Fig.\ 15, the two Bloch vectors rotate during the delay $\tau$ by different angles $\omega \tau$ around the $z$-axis.
In the most restricted case of UR pulses, the following UR($90^\circ_{-y}$) pulse brings all 
Bloch vectors from the equator into the $y$-$z$-plane (cf. Fig.\ 15 C). In contrast, the final Bloch vectors for PP$_0$ (Fig.\ 15 B) and PP$_R$,
ST$_R$ and s$^2$-COOP$_R$ pulses (Fig.\ 15 C) are not required to be located in this plane. However, the conditions for the Euler angles summarized in Table 3
ensure that for each offset frequency $\omega$, the $z$-component of the final Bloch vector corresponds to the value defined by the
target fringe pattern of Eq.\ \eqref{Mztarget}.
Hence for these pulse types, for each offset the final Bloch vector is only required to be located on a cone around the $z$-axis. In Fig.\ 15, the projections of the Bloch vectors onto the $z$-axis are indicated by grey triangles.

The presented optimal-control based approach for the efficient optimization of s$^2$-COOP pulses
can be generalized in a straightforward way to take into account additional aspects of practical interest
such as restrictions on the power or total energy of the control pulses (cf.\ \cite{Power-BEBOP}),
effects of amplitude and phase transients \cite{ESR_OC}
or the effects of relaxation during the pulses \cite{RC_BEBOP}.
In systematic studies of broadband UR \cite{UR_limits}, PP$_0$ \cite{BEBOP_limits} and 
PP$_R$ \cite{Iceberg} pulses,
it was found that for a desired value of the quality factor, %
the minimum pulse duration $T$ scales roughly linearly 
with the bandwidth $\Delta \nu$ and a similar scaling behavior is expected for 
s$^2$-COOP$_R$ pulses.

In addition to standard Ramsey experiments and e.g. the precise measurement of magnetic fields for a large range of field amplitudes
\cite{Wrachtrup},
potential applications of ST$_R$ and s$^2$-COOP$_R$ Ramsey pulses include stimulated echoes 
as well as two-dimensional spectroscopy.
The specific optimization parameters of the challenging test case considered here
was motivated by  2D-$^{13}$C-$^{13}$C-NOESY experiments
at future spectrometers
with ultra-high magnetic field strengths that are currently under development.
However, the presented algorithms can of course be used to optimize the performance of
Ramsey-type pulse sequence elements for any desired set of experimental parameters.
For example, significant gains compared to conventional approaches are
already expected if s$^2$-COOP pulses are designed for bandwidths corresponding to currently available
field strengths. 
In addition to the frequency labeling blocks of 2D-NOESY experiments discussed here,
in NMR spectroscopy the novel  Ramsey  90$^\circ$-$\tau$-90$^\circ$ building blocks
can be directly applied
in many other
2D experiments, such  as 2D exchange spectroscopy \cite{Ernst, Jeener1979}
and
also
in heteronuclear correlation experiments
such as heteronuclear single quantum coherence spectroscopy (HSQC) \cite{HSQC}  and 
heteronuclear multiple quantum coherence spectroscopy (HMQC) \cite{HMQC}. %
For example, s$^2$-COOP Ramsey pulses can be 
used as initial and final pulses in a modified INEPT block 
\cite{INEPT},
where 
instead of the standard 
central 180$^\circ$ refocusing pulses two  inversion pulses are applied to the spins of both nuclei  \cite{Levitt_Freeman_1980, Conolly_1991, INEPT_D_SE}, provided that the phase of the second Ramsey pulse is shifted by 90$^\circ$.

Beyond the
Ramsey scheme, which was considered here merely as an illustrative example,
it is expected that s$^2$-COOP pulses will find numerous applications in the control of complex quantum systems in spectroscopy, imaging and quantum information processing. %
In particular, the presented approach for the optimization of  s$^2$-COOP pulses can be used for the efficient and robust control schemes of general quantum systems and is 
not limited to the control of spin systems.

 \section*{Acknowledgements}
\footnotesize{
S.J.G. acknowledges support from
the DFG (GL 203/7-1) and SFB 631. M.B. thanks the Fonds der Chemischen
Industrie for a Chemiefonds
stipend.
}

\vfill \eject

\section{Appendix}

\subsection{Derivation of the relations between the Euler angles of  $S$ and $S^\prime$}

 In Table 1, the  relations between the offset-dependent Euler angles of  pulses $S$ and $S^\prime\in \{S^{-1},\ S_{ps},\ S_{ip}, \ S^{tr},\ S^{tr}_{ps}, S^{tr}_{ip}\}$ were summarized. 
Here it is shown explicitly how these relations can be derived from
well known properties of rotation operators
\cite{Euler3,Levitt_1986,URconstruction,UR_limits}.

\smallskip
For a given pulse $S$, the corresponding offset-dependent rotation operator 
${\bf S}(\omega)$ can be expressed in the ZYZ convention \cite{Euler3} in terms of the offset-dependent Euler angles $\gamma(\omega)$, $\beta(\omega)$ and $\alpha(\omega)$ as
\begin{equation}
\label{S_Euler}
{\bf S}(\omega)={\bf R}_z\{\alpha(\omega)\} {\bf R}_y\{\beta(\omega)\} {\bf R}_z\{\gamma(\omega)\},
\end{equation}
i.e.\ as a sequence of rotation operators ${\bf R}_a\{\varphi\}$ with rotation axis $a$ and rotation angle $\varphi$. As usual, the rotations operators are  ordered from right to left, i.e.\ in Eq. \eqref{S_Euler}
${\bf R}_z\{\gamma(\omega)\}$ is applied first, followed by 
$ {\bf R}_y\{\beta(\omega)\} $ and ${\bf R}_z\{\alpha(\omega)\}$.

\subsection{$S^\prime=S^{-1}$} 

\begin{eqnarray}
\label{S_prim_ipabcinv}
{\bf S}^{-1}(\omega)&=&\big({\bf R}_z\{\alpha(\omega)\} \ {\bf R}_y\{\beta(\omega)\} \ {\bf R}_z\{\gamma(\omega)\}\big)^{-1}
 \nonumber 
 \\
 &=&  {\bf R}^{-1}_z\{\gamma(\omega)\} \   {\bf R}^{-1}_y\{\beta(\omega)\} \  {\bf R}^{-1}_z\{\alpha(\omega)\}        
 \nonumber 
 \\
  &=&  {\bf R}_z\{-\gamma(\omega)\} \   {\bf R}_y\{-\beta(\omega)\} \  {\bf R}_z\{-\alpha(\omega)\}.
\end{eqnarray}
In the second line we used the standard relation $(ABC)^{-1}=C^{-1} B^{-1} A^{-1}$ and
in the third line we used the fact that the inverse of a rotation operator is simply obtained by inverting the sign of the rotation angle.

\subsection{$S^\prime=S_{ps}$} 

According to Table 1, for $S^\prime=S_{ps}$ the  pulse amplitude $u_{ps}(t)$
and pulse phase $\xi_{ps}(t)$ is given by
\begin{equation}
\label{uprimeps}
u_{ps}(t)=u(t), \ \ \ {\rm and} \ \ \ \xi_{ps}(t)=\xi(t+\pi).
\end{equation}
This corresponds to a $180^\circ$ rotation of the overall rotation operator ${\bf S}(\omega)$ around the $z$ axis:
\begin{eqnarray}
\label{S_prim_ps0}
{\bf S}_{ps}(\omega)&=&{\bf R}_z\{\pi\} {\bf S}(\omega) {\bf R}^{-1}_z\{\pi\}.
\end{eqnarray}
Inserting Eq.\ \eqref{S_Euler} in Eq.\ \eqref{S_prim_ps0}, we find
\begin{eqnarray}
\label{S_prim_ps}
{\bf S}_{ps}(\omega)&=&{\bf R}_z\{\pi\} \ {\color{blue} {\bf R}_z\{\alpha(\omega)\} {\bf R}_y\{\beta(\omega)\}  {\bf R}_z\{\gamma(\omega)\} }\ {\bf R}^{-1}_z\{\pi\} \nonumber \\
&=&{\color{blue} {\bf R}_z\{\alpha(\omega)\}} \ {\bf R}_z\{\pi\}  {\color{blue}{\bf R}_y\{\beta(\omega)\}}  {\bf R}^{-1}_z\{\pi\}\  {\color{blue}{\bf R}_z\{\gamma(\omega)\}} \nonumber \\
&=& {\bf R}_z\{\alpha(\omega)\} \ {\color{blue}\big({\bf R}_z\{\pi\}  {\bf R}_y\{\beta(\omega)\}  {\bf R}^{-1}_z\{\pi\}\big)}\  {\bf R}_z\{\gamma(\omega)\} \nonumber \\
&=&{\bf R}_z\{\alpha(\omega)\} \  {\color{blue}{\bf R}_{y}\{-\beta(\omega)\}} \  {\bf R}_z\{\gamma(\omega)\},
\end{eqnarray}
where the color is used to emphasize the individual transformations in each line.
In the second line we used the fact that rotations around the same axis commute and 
in the third line the terms are simply recolored to emphasize the relevant grouping for the next transformation.
The well known
relation
\begin{equation}
\label{formule1}
{\bf R}_a\{\pi\}  {\bf R}_b\{\varphi\}  {\bf R}^{-1}_a\{\pi\}={\bf R}_{-b}\{\varphi\}={\bf R}_{b}\{-\varphi\},
\end{equation}
which holds for arbitrary {\it orthogonal} rotation axes $a$ and $b$ and arbitrary angles $\varphi$ 
\cite{URconstruction, UR_limits},
was used in the fourth line 
with $a=z$, $b=y$ and $\varphi=\beta(\omega)$.

\subsection{$S^\prime=S_{ip}$}

According to Table 1, for $S^\prime=S_{ip}$ the  pulse amplitude $u_{ip}(t)$
and phase $\xi_{ip}(t)$ is given by
\begin{equation}
\label{uprimeps}
u_{ip}(t)=u(t), \ \ \ {\rm and} \ \ \ \xi_{ip}(t)=- \xi(t).
\end{equation}
As shown in \cite{Levitt_1986, URconstruction}, ${\bf S}_{ip}(\omega)$ is related to
${\bf S}(-\omega)$ via
\begin{eqnarray}
\label{S_prim_ip_Levitt}
{\bf S}_{ip}(\omega)&=&{\bf R}_x\{\pi\} {\bf S}(-\omega) {\bf R}^{-1}_x\{\pi\}.
\end{eqnarray}
 Replacing $\omega$ by $-\omega$ in Eq.\ \eqref{S_Euler}, ${\bf S}_{ip}(\omega)$ can be expressed in the form
\begin{eqnarray}
\label{S_prim_ipabc}
 {\bf S}_{ip}(\omega)&=& {\bf R}_x\{\pi\} \ \ {\color{blue} {\bf R}_z\{\alpha(-\omega)\} {\bf R}_y\{\beta(-\omega)\} {\bf R}_z\{\gamma(-\omega)\}}\ \ {\bf R}^{-1}_x\{\pi\}   
 \nonumber 
 \\
&=&{\bf R}_x\{\pi\}  {\bf R}_z\{\alpha(-\omega)\}  \ \ {\color{blue} {\bf R}^{-1}_x\{\pi\}{\bf R}_x\{\pi\}} \ \ 
{\bf R}_y\{\beta(-\omega)\}   {\bf R}_z\{\gamma(-\omega)\}  {\bf R}^{-1}_x\{\pi\}
 \nonumber  \\
&=&{\color{blue}\big({\bf R}_x\{\pi\}  {\bf R}_z\{\alpha(-\omega)\}  {\bf R}^{-1}_x\{\pi\}\big)}\ \  {\bf R}_x\{\pi\} 
{\bf R}_y\{\beta(-\omega)\}   {\bf R}_z\{\gamma(-\omega)\}  {\bf R}^{-1}_x\{\pi\}
 \nonumber \\
&=& {\color{blue}{\bf R}_{z}\{-\alpha(-\omega)\}}  \ \ {\bf R}_x\{\pi\} 
{\bf R}_y\{\beta(-\omega)\}  {\bf R}_z\{\gamma(-\omega)\}  {\bf R}^{-1}_x\{\pi\}
 \nonumber \\
&=& {\bf R}_{z}\{-\alpha(-\omega)\} \ \  {\bf R}_x\{\pi\} 
{\bf R}_y\{\beta(-\omega)\}  \ \ {\color{blue}{\bf R}^{-1}_x\{\pi\}{\bf R}_x\{\pi\}} \ \  {\bf R}_z\{\gamma(-\omega)\}  {\bf R}^{-1}_x\{\pi\}
 \nonumber \\
&=& {\bf R}_{z}\{-\alpha(-\omega)\} \ \  {\bf R}_x\{\pi\} 
{\bf R}_y\{\beta(-\omega)\}  {\bf R}^{-1}_x\{\pi\}  \ \  {\color{blue}\big({\bf R}_x\{\pi\} {\bf R}_z\{\gamma(-\omega)\}  {\bf R}^{-1}_x\{\pi\}\big)}
 \nonumber \\
&=& {\bf R}_{z}\{-\alpha(-\omega)\} \ \  {\bf R}_x\{\pi\} 
{\bf R}_y\{\beta(-\omega)\}  {\bf R}^{-1}_x\{\pi\}  \ \   {\color{blue}{\bf R}_{z}\{-\gamma(-\omega)\} }
 \nonumber \\
&=& {\bf R}_{z}\{-\alpha(-\omega)\} \ \ {\color{blue}\big( {\bf R}_x\{\pi\} 
{\bf R}_y\{\beta(-\omega)\}  {\bf R}^{-1}_x\{\pi\}\big)}  \ \   {\bf R}_{z}\{-\gamma(-\omega)\} 
 \nonumber \\
&=& {\bf R}_{z}\{-\alpha(-\omega)\} \ \  {\color{blue}
{\bf R}_{y}\{-\beta(-\omega)\} }  \ \   {\bf R}_{z}\{-\gamma(-\omega)\}.
\end{eqnarray}
In the second and fifth line, the identity operator ${\bf R}^{-1}_x\{\pi\}{\bf R}_x\{\pi\}={\bf 1}$ was inserted.
Eq.\ \eqref{formule1} was used in line 4
with $a=x$, $b=z$, $\varphi=\alpha(-\omega)$,
in line 7
with $a=x$, $b=z$, $\varphi=\gamma(-\omega)$
and in line 9
with $a=x$, $b=y$, $\varphi=\beta(-\omega)$.

\subsection{$S^\prime=S^{tr}$}

According to Table 1, for $S^\prime=S^{tr}$ the  pulse amplitude $u^{tr}(t)$
and phase $\xi^{tr}(t)$ is given by
\begin{equation}
\label{uprimeps3}
u^{tr}(t)=u(T-t), \ \ \ {\rm and} \ \ \ \xi^{tr}(t)=\xi(T-t).
\end{equation}
As shown in \cite{Levitt_1986, URconstruction}, ${\bf S}^{tr}(\omega)$ is related to
${\bf S}(-\omega)$ via
\begin{eqnarray}
\label{S_prim_tr_Levitt}
{\bf S}^{tr}(\omega)&=&{\bf R}_z\{\pi\} {\bf S}^{-1}(-\omega) {\bf R}^{-1}_z\{\pi\}.
\end{eqnarray}
Based on Eq.\ \eqref{S_prim_ipabcinv}, 
\begin{eqnarray}
\label{S_prim_ipabcg}
{\bf S}^{-1}(-\omega)&=&  {\bf R}_z\{-\gamma(-\omega)\} \   {\bf R}_y\{-\beta(-\omega)\} \  {\bf R}_z\{-\alpha(-\omega)\}        
\end{eqnarray}
which can be inserted in Eq. \eqref{S_prim_tr_Levitt}:
\begin{eqnarray}
\label{S_prim_tr_Levitt1}
{\bf S}^{tr}(\omega)&=&{\bf R}_z\{\pi\} \ \    {\color{blue}{\bf R}_z\{-\gamma(-\omega)\} \   {\bf R}_y\{-\beta(-\omega)\} \  {\bf R}_z\{-\alpha(-\omega)\} }  \ \  {\bf R}^{-1}_z\{\pi\}  \nonumber \\
&=&    {\color{blue} {\bf R}_z\{-\gamma(-\omega)\}} \  {\bf R}_z\{\pi\} {\color{blue} {\bf R}_y\{-\beta(-\omega)\} }{\bf R}^{-1}_z\{\pi\}  \  {\color{blue}{\bf R}_z\{-\alpha(-\omega)\}}   \nonumber \\
&=&   {\bf R}_z\{-\gamma(-\omega)\} \   {\color{blue} \big({\bf R}_z\{\pi\}{\bf R}_y\{-\beta(-\omega)\} {\bf R}^{-1}_z\{\pi\} \big)} \  {\bf R}_z\{-\alpha(-\omega)\}  \nonumber \\
&=&    {\bf R}_z\{-\gamma(-\omega)\} \   {\color{blue}  {\bf R}_{y}\{\beta(-\omega)\} }\  {\bf R}_z\{-\alpha(-\omega)\}.  
\end{eqnarray}
In the first line the $z$-rotations commute and
Eq.\ \eqref{formule1} was used in line 4 
with $a=x$, $b=y$, $\varphi=-\beta(-\omega)$.

\subsection{$S^\prime=S^{tr}_{ps}$} 

The Euler rotations of ${\bf S}^{tr}_{ps}=({\bf S}^{tr})_{ps}=({\bf S}_{ps})^{tr}$ are obtained based on
Eqs.\ \eqref{S_prim_ps} and \eqref{S_prim_tr_Levitt1}:
\begin{eqnarray}
\label{S_prim_trps}
{\bf S}^{tr}_{ps}(\omega)&=&\big({\bf S}_{ps}(\omega)\big)^{tr}
 \nonumber 
 \\
&=&\big( {\color{blue} {\bf R}_z\{\alpha(\omega)\} \  {\bf R}_{y}\{-\beta(\omega)\} \  {\bf R}_z\{\gamma(\omega)\}}\big)^{tr}
 \nonumber 
 \\   
&=&{\color{blue} {\bf R}_z\{-\gamma(-\omega)\}  \  {\bf R}_{y}\{-\beta(-\omega)\} \    {\bf R}_z\{-\alpha(-\omega)\}}.  \nonumber 
\end{eqnarray}

\subsection{$S^\prime=S^{tr}_{ip}$} 

The Euler rotations of ${\bf S}^{tr}_{ip}=({\bf S}^{tr})_{ip}=({\bf S}_{ip})^{tr}$ obtained based on
Eqs.\ \eqref{S_prim_ipabc} and \eqref{S_prim_tr_Levitt1}:
\begin{eqnarray}
\label{S_prim_trps4}
{\bf S}^{tr}_{ip}(\omega)&=&\big({\bf S}_{ip}(\omega)\big)^{tr}
 \nonumber 
 \\
&=&\big( {\color{blue} {\bf R}_{z}\{-\alpha(-\omega)\} \ \  
{\bf R}_{y}\{-\beta(-\omega)\}   \ \   {\bf R}_{z}\{-\gamma(-\omega)\}} \big)^{tr}
 \nonumber 
 \\   
&=&{\color{blue} {\bf R}_z\{\gamma(\omega)\}  \  {\bf R}_{y}\{-\beta(\omega)\} \    {\bf R}_z\{\alpha(\omega)\}}.  \nonumber
\end{eqnarray}

\vfill eject

\section*{References}

\end{document}